\documentclass[12pt]{article}
\usepackage{graphicx,epsfig,amsmath,amssymb,pstricks}

\begin{document}

\title{\bf Building the Observer into the System:\\
Toward a Realistic Description of Human Interaction with the World}

\author{{Chris Fields}\\ \\
{\it 243 West Spain Street}\\
{\it Sonoma, CA 95476 USA}\\ \\
{fieldsres@gmail.com}}
\maketitle

\begin{abstract}
Human beings do not observe the world from the outside, but rather are fully embedded in it.  The sciences, however, often give the observer both a ``god's eye'' perspective and substantial \emph{a~priori} knowledge.  Motivated by W. Ross Ashby's statement, ``the theory of the Black Box is merely the theory of real objects or systems, when close attention is given to the question, relating object and observer, about what information comes from the object, and how it is obtained'' ({\em Introduction to Cybernetics}, 1956, p. 110), I develop here an alternate picture of the world as a black box to which the observer is coupled.  Within this framework I prove purely-classical analogs of the ``no-go'' theorems of quantum theory.  Focussing on the question of identifying macroscopic objects, such as laboratory apparatus or even other observers, I show that the standard quantum formalism of superposition is required to adequately represent the classical information that an observer can obtain.  I relate these results to supporting considerations from evolutionary biology, cognitive and developmental psychology, and artificial intelligence.

\end{abstract} 



\begin{quote}
The theory of the Black Box is merely the theory of real objects or systems, when close attention is given to the question, relating object and observer, about what information comes from the object, and how it is obtained.
\begin{flushright}
--- W. Ross Ashby, 1956 (\cite{ashby:56}, p. 110)
\end{flushright}
\end{quote}

\section{Introduction} 

Modern science is built on two far-reaching ideas: that there is a way that the world works, and~that we human beings are not separate from but are rather part of the world.  The first idea matured in the 17th and 18th centuries, reaching its full expression---but by no means widespread acceptance---in Laplace's concept of a mechanical, fully-deterministic universe.  The maturation of the second idea did not begin until the birth of modern biology and psychology in the mid-19th century, and it is not yet complete.  While both biology and psychology firmly place human beings and other organisms within the world, and while the enormous progress made in these disciplines during the past 50 years allows the construction of increasingly-sophisticated models of human beings and other organisms as systems that acquire information from their environments and use that information to act on their environments, ``the observer'' and her alter ego, ``the experimenter'' still stand outside of our most basic physical theories.  

The fundamental fact required to incorporate the observer/experimenter (hereafter, in keeping with tradition, simply ``the observer'') into physical theory has, however, been recognized for well over a century.  It was first formally elucidated by Boltzmann in the 1880s: reducing uncertainty by acquiring information requires the expenditure of energy.  The minimal energetic cost of acquiring and recording one bit, i.e., the answer to one yes/no question, is known: it is $ln2~kT$, where $k$ is Boltzmann's constant and $T$ is temperature \cite{landauer:61, landauer:99, bennett:03}.  However, $k$ is small, $ k \sim 1.38 \times 10^{-23}$ J/K in SI units, so at physiological temperature, $T$ = 310 K, the minimal per-bit energetic cost $ln2~kT \sim 3 \times 10^{-21}$ J is vastly smaller than typical macroscopic energies.  This tiny value for the per-bit energetic cost of observation justified the traditional ideal, implicit in classical physics, of the ``detached'' observer whose observational activities had no impact on the observed world.  

When, almost 50 years after Boltzmann's discovery, the early development of quantum theory introduced a formalism in which observation \textit{did} impact the world---collapsing wave functions and even determining whether entities would behave as waves or particles---this violation of the ideal of detachment quickly assumed metaphysical proportions.  From the remarkable Solvay Conference of 1927 \cite{baccia:06} onwards, debates about the physical meaning of the quantum formalism have remained largely metaphysical (for synoptic reviews, see~\cite{landsman:07, wallace:08}; for recent surveys of interpretative positions, see~\cite{norson:13, schloss:13, sommer:13}).  Physicists regularly deplore this situation, with Fuchs, for example, likening interpretative stances to religions \cite{fuchs:10} and Cabello labelling the entire interpretative landscape a ``map of madness''~\cite{cabello:15}.  Nonetheless, substantial numbers of physicists---perhaps most famously, Bell \cite{bell:90}---maintain that the observer and the process of observation can play no role in any theory, and in particular no role in any \textit{physical} theory, describing a world that worked in whatever way it worked long before humans existed and will continue to work in that way long after humans are gone.  Even physicists who embrace some form of what Fuchs \cite{fuchs:16} has called, following Wheeler \cite{wheeler:83}, ``participatory realism'' may insist, as~Fuchs himself has done, that physics can provide no \textit{theory} of the observer \cite{fuchs:10}.  If physics can offer no theory of the observer, however, the observer still stands outside of physics, whether as a ``participant'' or not.  

We \textit{know}, however, what observers do.  Observers acquire information and, in their complementary role as actors on their environments, they create information.  Biology and psychology tell us, in ever increasing detail, \textit{how} they acquire information and \textit{how} they act to create more of it.  Hence observation, together with its complement, experimentation, is a process of information exchange between an observer and that observer's environment.  Both the observer and the observer's environment are parts of the world, and if the ``environment'' is expanded to become the observer's complement, the two together constitute the world.  Boltzmann's insight then tells us that any exchange of information between observer and environment involves an exchange of energy, and hence a \textit{physical interaction}, between observer and environment.  This simple consideration renders the detached observer a mere approximation, one that may be expected sometimes---and possibly always---to fail.    

That the idealization of the detached observer must be rejected even on classical-physics grounds~is, of course, not a new idea: it is a cornerstone of second-order cybernetics.  von Uexk\"{u}ll~\cite{vonUexkull:57}, von Foerster \cite{vonFoerster:76}, Kampis \cite{kampis:96}, Koenderink \cite{koenderink:14} and many others have argued for it explicitly.  R\"{o}ssler's concept of endophysics \cite{rossler:87} is based on it.  The critical question in any theoretical framework that rejects the detached observer is that of how the ``epistemic cut'' \cite{vonNeumann:55} separating the observer from the observed environment is defined, and in particular, whether it is defined in a way that truly places the observer within the world.  As Kampis puts it, ``what endophysics aspires for is the study of the encompassing big black box of which the observer and his epistemology are a part ... an enclosed observer is bound to the same laws as those of the system observed'' (\cite{kampis:96},  p. 265).  As von Neumann recognized \cite{vonNeumann:55}, the observer being ``bound by the same laws'' is assured if the position of the cut is arbitrary (\textit{cf.} Pattee, ``The cut itself is an epistemic necessity, not an ontological condition'' \cite{pattee:01}, p. 13).  If the observer is ``special'' in some way, so that the observer-environment boundary cannot be ``erased'' within the theory without affecting the behavior of the world---as it appears to be the case, for example, in the endophysical framework of Kauffman and Gare \cite{kauffman:15}---then the observer remains in an important sense outside of the theory.

Classical cybernetics provides us with a simple formal model of the information exchange between the observer and the observed environment: the theory of the Black Box.  My goal in this paper is to take seriously both the Black Box model and the requirement that the epistemic cut be arbitrary and to see where these assumptions jointly lead.  What is offered here is, therefore, not a new model of observation, but rather an exploration in some depth of an old model of observation conjoined with an even-older general constraint on model construction.  After reviewing the Black Box model and the associated epistemic cut in the next section, I show that conjoining the Black Box model with an arbitrarily-movable cut leads to ``no-go'' theorems, i.e., theorems limiting the inferences that may be drawn from either theoretical considerations or observational outcomes, that are classical analogs of no-go theorems familiar from quantum information theory.  I then show that if a classical black-box world is assumed, by hypothesis, to comprise a collection of bounded, causally-independent ``objects'',  finite observations can only identify superpositions of such objects.  The classical Black Box model with an arbitrary epistemic cut thus requires us to view even a classical world in quantum-theoretic terms.  The traditional ``classical worldview'' of bounded, independently-observable and independently-manipulable objects only emerges when we ignore the constraints of the Black Box model and assume that observers have a priori knowledge and local causal power that the Black Box model---and indeed, the formalisms of both quantum and classical physics---tell us that observers cannot obtain.  To place this result in a larger context, I briefly discuss ideas and observations from evolutionary biology, cognitive and developmental psychology, computer science and artificial intelligence that point toward this same conclusion.  I close by suggesting that if observers and observation are placed firmly within the world, its theory of the world is the only theory of the observer that physics needs.  The observer only steps out of physical theory---and indeed, out of the physical world---when physics contradicts itself, granting to the observer knowledge and causal power that physics itself tells us no worldly observer can have.

\section{The Black Box Model}

The Black Box model is a formal representation of the interaction between an observer and the system being observed.  This section reviews the informal notion of a ``system'' with particular emphasis on the assumption of separability.  It then presents the Black Box model and provides reasons for believing that it accurately describes the interactions between observers and physical systems.  It~briefly discusses the relation between the sense of observer dependence entailed by the Black Box model and the informal notion of ``subjectivity'' with which it is sometimes associated.

\subsection{Systems and Separability}\label{sec2.1}

The terms `universe' and `world' are here used interchangeably to denote the maximal closed system containing all possible observers and all possible observed systems as components.  Here `closed' has the usual meaning of not interacting with any other system.  As observation requires interaction, any closed system is unobservable; the universe, as a closed system, is therefore unobservable.  Setting~special relativity aside for simplicity, the `observed world' or `environment' of any observer can be regarded as the complement of that observer within the world.  In this case, given any arbitrary choice of observer, that observer together with that observer's environment jointly compose the world (taking special relativity into account converts this complement into a light cone).

The terms `system' and `physical system' are here used interchangeably to denote a component of the world, \textit{not} a mathematical or other formal model of such a component as is sometimes intended by other authors.  A ``system'' in the latter, formal-model sense can be stipulated.  A system in the sense employed here can only be observed and/or acted upon, after which a model of that system may be constructed or stipulated.  The term `component' is here used informally; it can be made precise by stipulating a physical model of the universe which is then decomposed.  If the universe is represented by a classical configuration (or phase) space, decomposition is implemented by the Cartesian product operator $\times$, while if the universe is represented by a Hilbert space, decomposition is implemented by the tensor product operator $\otimes$.  In either case, the factors into which the universe is decomposed represent systems as defined here.  The Black Box model is a model of the interaction between observers and systems, i.e., a model of the interaction between two components of the world, an observer (which is always itself a system) and a system (which may itself also qualify as an observer).  

The usage employed here may be compared to that recently employed by Kitto \cite{kitto:14}.  It is consistent with the idea that a system is ``a set of entities that are interacting via a set of relationships'' (\cite{kitto:14}, p. 542), provided that the entities and relationships referred to are components and characteristics, respectively, of the world, as opposed to models, approximations, or simplifications of such components or characteristics.  The current usage differs, however, from that of Kitto in that it \textit{does not} assume that a system is ``separate from its surroundings'' (\cite{kitto:14}, p. 541).  In particular, it does not assume that a system is \textit{separable} from its surroundings as this term is employed in physics, i.e., it does not assume that a system occupies or can be assigned a state independently of the state occupied by or assigned to its surroundings.  As discussed in detail below, this latter provision allows the observer-system boundary to be moved arbitrarily without altering the composition or the behavior of the world, i.e., of the composite system comprising the observer plus the observed system.

Separability is an intuitively appealing and quite natural assumption; hence rejecting it from the outset can be regarded as radical.  Separability is, however, despite its intuitive appeal an \textit{assumption}, indeed (as will be shown below) a very strong one.  While failures of separability are typically identified (by definition) with entanglement in the quantum theory literature, moreover, they can arise in multiple formal and interpretative contexts.  Tipler, for example, has shown that requiring classical dynamics to be strictly deterministic reproduces the formalism of unitary quantum theory and hence entails non-separability \cite{tipler:14}; indeed he shows that the ``quantum potential'' required for strict classical determinism is identical to the ``guiding field'' of Bohmian quantum mechanics \cite{bohm:87}.  While Bohm interpreted this guiding field as non-local, however, Tipler interprets it as strictly local.  In a similar vein, Rosen \cite{rosen:86} formalizes the notion of a constraint hierarchy in classical complex systems proposed by Polanyi \cite{polanyi:68}, showing that it can be represented as an arbitrarily extended hierarchy of potential functions.  Rosen concludes, ``a complex system requires an \textit{infinite} mathematical object for its description ... it is quite clear that there is no such thing as a set of \textit{states}, assignable to such a system once and for all'' (\cite{rosen:86},  p. 190, emphasis in original).  Here again we have a failure of separability in a \textit{classical} context, one to which the quantum-theoretic concept of entanglement would ordinarily be thought irrelevant.

Separability is closely associated with contextuality, i.e., with the potential dependence of system behavior on how, in what setting, or with what preparations measurements are made.  Kitto~\cite{kitto:14} distinguishes three kinds of contextuality: dependencies of system components on each other, dependencies of system behavior on how it is measured, and dependencies of system behavior on the environmental setting.  The distinctions between these types of contextuality depend, however, on the assumption of separability.  If the components of a separable system are non-separable, the~first kind of contextuality appears as Kitto describes (\cite{kitto:14},  p. 551); similarly if a system $\mathbf{S}$ is not separable from the apparatus employed to measure its state, but the joint system $\mathbf{S}^{\prime}$ comprising $\mathbf{S}$ plus the apparatus is separable from its environment, the second kind of contextuality appears.  If, however, the assumption of separability is dropped altogether, all three kinds of contextuality can appear and they cannot be distinguished.  As shown in detail in \S 3 below, if a system cannot be bounded, its components cannot be identified; \textit{ipso facto} they cannot be distinguished from components of the apparatus, the environment, or for that matter, the observer.  Here the distinction between the \textit{identification} of components---something that must be accomplished by observation---and the \textit{stipulation} of components within an hypothesized model (and hence the fundamental distinction between systems and models) is clearly essential.  

By not assuming separability as an a priori condition on systems, the present analysis permits arbitrary contextuality.  As shown in \S 3 below, an arbitrarily-movable epistemic cut renders the possibility of such contextuality inescapable.  Potential effects of contextuality may in practice be~\textit{ignored}, i.e., simply assumed not to exist as they often are, but the consequences of doing so cannot be predicted and may be severe. 

\subsection{Definition of the Black Box}

A \textit{black box} as described by classical cybernetics \cite{ashby:56} is a system to which observers have only exterior access.  They can form hypotheses about what is going on inside the box, they can formulate and test by further observation models of what is going on inside the box, but they cannot disassemble the box, examine its interior workings, and determine by such observations the dynamics that the box~executes.

An observer interacts with a black box by sending it strings of bits; traditionally this interaction is considered to involve turning dials, pushing buttons, and so forth.  The black box, in turn, sends the observer strings of bits, e.g., by moving pointers on dials or changing the numbers in a digital display.  The observer-box interaction is considered to take place in episodes, i.e., the observer does something to the box, waits for a certain period for the box to respond, and then does something else.  The box may respond by doing nothing.  The observer may also do nothing, in hopes of observing spontaneous behavior on the part of the box.

The observer is assumed to be a finite system with finite energy resources.  The finite energetic cost of encoding each bit limits the observer to sending only a finite number of bits to the box, and~to recording only a finite number of bits from the box, in each of a finite number of episodes of interaction.  The lengths of these ``input'' and ``output'' bit strings can be considered the ``resolution'' of the inputs to and outputs from the box.  For simplicity, the input and output resolutions are considered to be identical.  In this case, the information exchange between observer and box can be considered a classical information channel \cite{shannon:48} with a fixed capacity equal to the resolution.  Again for simplicity, this classical information channel can be considered to be noise-free.  The consequences of relaxing the restriction to finite communication through this classical channel is discussed in \S 3.3 below.

Nothing prevents the observer from sending the same input to the box multiple times; similarly, nothing prevents the box from sending the same output to the observer multiple times.  If the channel capacity and hence the resolution of the inputs and outputs is small and the number of episodes of interaction is large, such repetitive behavior is inevitable.

The observer is assumed to be capable of observing her own inputs to the box, and to have a memory on which both the input sent and the output received in each episode are both written.  Because the number of interaction episodes is finite, this memory is finite.  It is assumed for simplicity that the observer's memory is sufficient to record all of the interaction episodes; this assumption is discussed further in \S 3.3 below.  The observer can be considered to consult the memory in order to frame hypotheses about the box's behavior and to freely choose, based on these hypotheses, what~input to send to the box in the next episode.  The meaning of ``freedom'' in this context is also discussed in \S 3.3 below.

These descriptions, characterizations and assumptions can be formalized with the following:

\textbf{Definition}:  A \textit{black box} is a system about which no observer can have more (non-hypothetical) information than is contained in a finite list of finite-length bit strings representing observed input-output transitions.

\begin{figure}
\centering
\includegraphics[width=16cm]{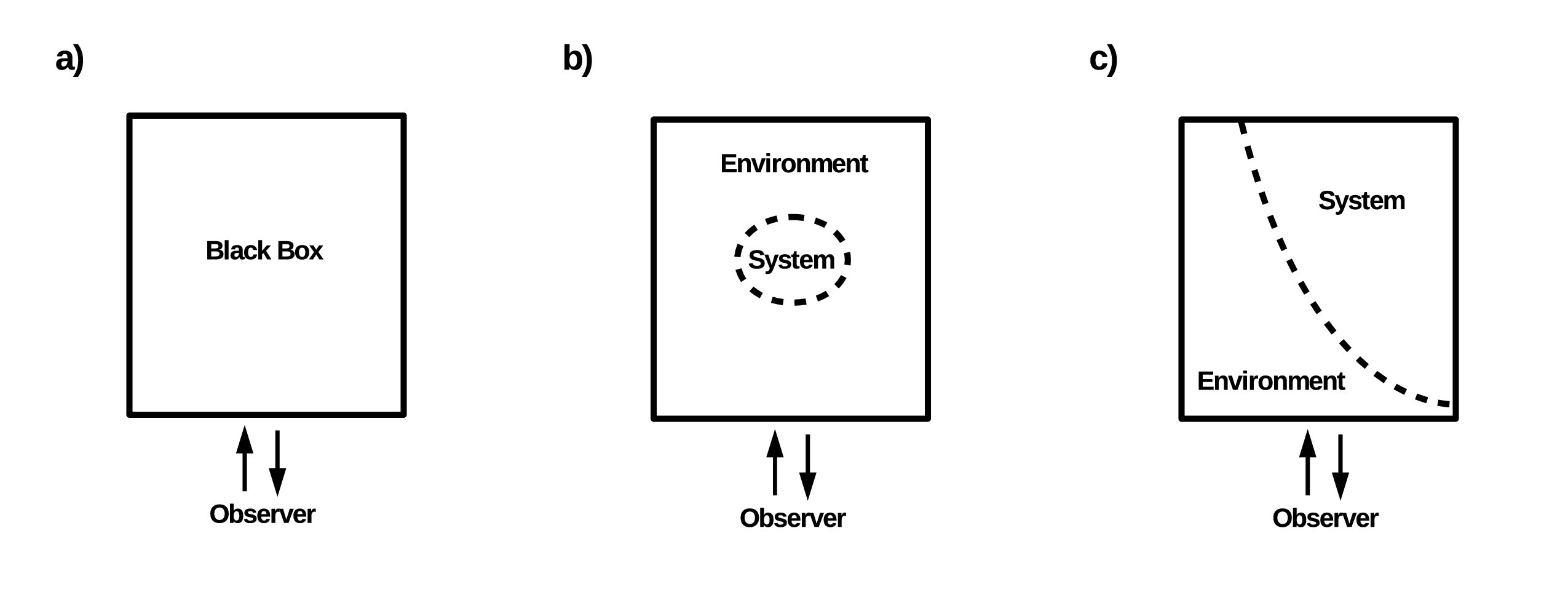}
\caption{The Black Box model and two related concepts.  (\textbf{a})  An observer interacts with a black box by exchanging information with the box through a classical channel of fixed, finite capacity; (\textbf{b})~The~``environment as witness'' formulation of decoherence permits observers to access information about a physical system only via the mediation of a surrounding environment;  (\textbf{c}) Decompositional equivalence requires an observer's interactions with a composite system to be independent of the placement of subsystem boundaries within that system.}
\label{fig1}
\end{figure}

The interaction between the observer and the box can be represented as in Figure \ref{fig1}a.  The~observer collects finite data about the behavior of the box ``from the outside'' and uses these data---the only data that are available---to understand and frame hypotheses about the box's internal structure and dynamics and to use these hypotheses to predict the output that will be produced by the box in response to each input.

\subsection{The Black Box as a Model of Physical Systems}

As indicated by the opening quotation, Ashby considered all ``real objects or systems'' to be black~boxes, or at least to become black boxes as their structures and behaviors were probed at finer and finer resolutions.  Opening the case of a laptop computer---or increasingly, even the engine compartment of an automobile---reveals within merely an arrangement of fully-encapsulated parts, ``components'' with internal structures hidden from the user.  Such components are black boxes by design, meant to be installed within the larger system without modification and to be fully interchangable with replacements having the same input-output behavior without affecting the performance of the larger system.  Further investigation of such components reveals that they too contain components.  Even when a component is easily recognizable and ``simple''---perhaps it is a machine screw or a wire---it is still possible to ask what internal structure explains its properties or behavior.  However, at some point such questioning stops, leaving whatever components that remain black boxes.

Probing any physical system deeply enough reveals structure and behavior only adequately describable using quantum theory.  It is here that the black-box nature of physical systems becomes~explicit.  As Bohr repeatedly emphasized, in making measurements of the behavior of a quantum system one is inevitably forced to make use of macroscopic instruments.  One must interact, in other words, with a macroscopic composite system comprising the quantum system of interest together with the apparatus used to measure its behavior.  No independent access to the quantum system is possible.  This lack of independent access renders the quantum system a black~box.  In~addition, the physical coupling of the quantum system of interest to the apparatus blurs the boundary between the two, a blurring that is formalized as entanglement by the unitary time evolution of the Schr\"{o}dinger equation.  As Bohr put it, ``it is no longer possible sharply to distinguish between the autonomous behavior of a physical object and its inevitable interaction with other bodies serving as measuring instruments, the direct consideration of which is excluded by the very nature of the concept of observation in itself'' (\cite{bohr:37},  p. 290). 

It is important here to emphasize that characterizing a quantum system---or any system---as a black box is making a statement about \textit{observational access} to that system.  It is straightforward to write down precise formal models of many quantum systems in either mixed or pure states and to claim, on~the basis of such a model, that a system is well understood.  Observing a model, however, is not the same thing as observing a system.  Arbitrarily precise models can be constructed for any black box, and they can accurately predict some finite number of observations or not.  The existence of such a~model, however, has no bearing on whether the system being modelled is a black box.  A system is a black box if and only if it satisfies the definition given above, i.e., if and only if it is a system about which no observer can have more (non-hypothetical) information than is contained in a finite list of finite-length bit strings representing observed input-output transitions.  The statements of Ashby and Bohr quoted above are statements about observational access, not statements about the possibility of model construction.

In the Copenhagen interpretation of Bohr and Heisenberg that became the standby of ``textbook'' quantum theory (e.g., \cite{landau:77}; see also \cite{mermin:89}), it is the amplification of signals indicating the presence of quantum phenomena by a directly coupled macroscopic apparatus that converts the fragile information contained in quantum states into the robust, recordable classical information of observational~outcomes.  The more modern, though by now almost 50 years old, version of this view is the theory of environmental decoherence \cite{zeh:70, zeh:73, zurek:81, zurek:82, joos-zeh:85, zurek:98, zurek:03, schloss:07}, in which exposure to and hence entanglement with any macroscopic environment, intentional or not, converts quantum information into classical information.  Ollivier, Poulin and Zurek introduced the ``environment as witness'' formulation of environmental decoherence \cite{zurek:04, zurek:05}, later elaborated as the theory of ``quantum Darwinism'' \cite{zurek:06, zurek:09, zurek:10, zurek:12, zurek:14} or ``quantum state information broadcasting'' \cite{chiribella:06, korbicz:14, horodecki:15, brandao:15} in recognition of the fact that observers typically obtain information about physical systems by interacting with the surrounding environment, typically the surrounding ambient photon field.  This more recent formulation of environmental decoherence makes it explicit that observers are \textit{not directly coupled} to the systems that their observational outcomes purport to describe, but are rather directly coupled to an intervening environment---effectively a very large apparatus---comprising vast numbers of unobserved degrees of freedom. 

Central to the environment as witness formulation and quantum Darwinism is the requirement that environmental encodings of quantum states be \textit{redundant} in the sense that many observers at different locations can each obtain the same information about the system of interest without disturbing either the interaction of the system with its environment or the information accessed by the other observers (cf. \cite{zurek:05}, Figure 1 and surrounding discussion).  This redundancy requirement prevents observers from accessing information about the system by any means \textit{other than} through the environment, as any such access would disturb the system and hence destroy the redundancy of the encoding \cite{fields:10, fields:11}.  The redundancy requirement, in other words, makes the environment a black box, a container that prevents direct observational access to any systems contained in its interior (Fig. 1b).  If observations of quantum systems satisfy the constraints imposed by the environment as witness formulation of decoherence, therefore, they can be characterized as interactions with a black box.

The theoretical framework of environmental decoherence assumes ``no-collapse'' or unitary quantum theory, i.e., quantum theory in which there is no \textit{physical} process of quantum state collapse and hence no \textit{physical} violation of unitarity \cite{schloss:06}.  Unitary quantum theory applies, however, only to closed systems.  As first explicitly recognized by Everett \cite{everett:57}, this restriction of unitary evolution to closed systems has two important consequences: (1) the universe, being a closed system, must~evolve unitarily; and (2) the universe can contain no closed systems, as the absence of interactions between such systems and their surroundings would permit violations of unitarity.  The environmental decoherence framework is explicitly designed to cope with the openness of all possible systems to interaction---and hence to observation---that is implied by point \# 2.  Schlosshauer, for example, remarks when introducing the decoherence framework that ``over the past three or so decades it has been slowly realized that the isolated-system assumption---which, as we have described above, had proved so useful in classical physics and had simply been taken over to quantum physics---had in fact been the crucial obstacle to an understanding of the quantum-to-classical transition'' (\cite{schloss:07},  p. 3--4).

If unitary quantum theory is correct, then there are no \textit{physically} classical systems, only apparently or ``for all practical purposes'' (FAPP; terminology due to Bell \cite{bell:90}) classical systems.  Alternative~theories in which unitarity is physically violated have been proposed, and are sometimes (incorrectly) regarded as ``interpretations'' of quantum theory as opposed to alternative theories; examples include the theories of Ghirardi, Rimini and Weber \cite{ghirardi:86}, Penrose \cite{penrose:96} or Weinberg \cite{weinberg:12}.  In such theories, there are physically classical systems.  However, these theories have at present no empirical support (e.g., \cite{schloss:06, jordan:10, wiseman:15}) and are increasingly challenged by theoretical approaches such as holography that invoke quantum theory at very large scales (e.g., \cite{swingle:12, saini:15, susskind:16}) and by derivations of quantum theory from purely information-theoretic foundations (see Section \ref{sec3.4} below).  This situation presents at least a \textit{prima facie} challenge to any theoretical approach that assumes or implies the existence of physically-classical systems.  Kitto, for example, remarks that in at least some cases ``measurement merely records reality; it does not influence what is found during that measurement'' while cautioning that ``such assumptions, while sometimes correct, can be markedly dangerous'' (\cite{kitto:14},  p. 547).  If unitary quantum theory is correct, such assumptions maybe \textit{useful} in some circumstances, but they are never correct in the strong sense of accurately describing reality.  If the universe indeed exhibits large-scale entanglement as unitary quantum theory would imply, regarding ``the notion of separability to be key to the definition of a system'' (\cite{kitto:14},  p. 542) results in a universe without systems.

\subsection{The Epistemic Cut and Decompositional Equivalence}

The universality of the Black Box model for physical systems is, however, also supported by theoretical considerations far more fundamental than decoherence.  As noted earlier, both classical and quantum physics represent the states of physical systems as vectors in state spaces, the real configuration (or phase) space in classical physics and the complex Hilbert space in quantum physics.  The product operators that implement state-space factorization and its inverse, state-space combination are in each case associative, i.e., $A \times B \times C = (A \times B) \times C = A \times (B \times C)$ in classical physics and $A \otimes B \otimes C = (A \otimes B) \otimes C = A \otimes (B \otimes C)$ in quantum physics.  Both theories, moreover, represent~interactions between systems by linear operators, specifically Hamiltonians.  These operators are additive, i.e., $H_{(A \times B)} = H_{A} + H_{B} + H_{AB}$ in classical physics and $H_{(A \otimes B)} = H_{A} + H_{B} + H_{AB}$ in quantum physics, where in each case the final term $H_{AB}$ represents the $A - B$ interaction.  In~both~theories, therefore, \textit{decomposing a system into two or more components makes no difference to the composite system's structure or behavior}.  The epistemic cut between components is, in this case, purely~epistemic as Pattee \cite{pattee:01} requires; where it is placed makes no difference to the composite system.  An observer interacting with the composite system cannot tell, either by characterizing its state space or by measuring its dynamics, that it has been decomposed into components.  Both the boundaries between the internal components and the interactions defined at these internal boundaries are invisible to an outside observer.  The composite system is, from the outside observer's perspective, a black box within which its components are fully contained (Figure \ref{fig1}c).  In describing quantum Bayesianism, Fuchs employs phenomenological language to make this point, referring to the ``hermetically sealed'' character of a black-box world as ``interiority'' and attributing an inviolable interiority to all systems, large and small (\cite{fuchs:10},  p. 20).

The invisibility of internal component boundaries to outside observers, previously termed \textit{decompositional equivalence} \cite{fields:12a}, has significant consequences for decoherence and particularly for quantum Darwinism \cite{fields:12b, fields:14a, fields:16a}.  If the classical information acquired by outside observers of a composite system does not depend on the placement of subsystem boundaries within that system, then it cannot depend on interactions defined at those boundaries.  Observers of an environment cannot, therefore, determine the boundaries of systems embedded within that environment given only classical information encoded by that environment.  Any such encoding is arbitrarily ambiguous about the boundaries of embedded systems.  Multiple observers can interact with the same, redundant~encoding, hypothesize radically different system-environment boundaries, and infer radically different system states as a result.  They can agree completely about the observational outcomes they have~obtained, but disagree about both the system and the system-environment interaction that produced them.  This~principled ambiguity about the states of internal systems extends even to whether they are~coherent, i.e., entangled, or environmentally decohered, a phenomenon termed ``entanglement relativity'' \cite{zanardi:01, zanardi:04, dugic:06, dugic:08, torre:10, harshman:11, thirring:11, dugic:12}.

The goal of observation is to determine the state space and the dynamics, i.e., the self-Hamiltonian, of the system being observed.  The classical Black Box model, the environment as witness formulation of decoherence, and the formal constraints imposed on either quantum or classical physics by decompositional equivalence all locate the physical interaction that implements observation at a single boundary: the boundary between the observer and the observed composite environment, i.e., between the observer and the entire rest of that observer's observable universe.  The physical interaction at this boundary implements the classical information channel between the observer and the composite environment that is that observer's observed world.  Only information that flows through this classical channel contributes to observational outcomes.  The observer can hypothesize, on the basis of these observational outcomes, the existence of internal boundaries within the composite environment and hence internal structure and interactions within the observed world.  These hypotheses can only be tested, however, by examining their consequences for the observed world as a composite whole, because the observed world as a whole is the only system with which the observer directly interacts.

\subsection{Observer Dependence and ``Subjectivity''}

A Black Box model of the observer-environment interaction entails a precise sense of observer dependence: each decomposition of the universe into an observer and that observer's environment is~unique, so each observer observes a unique environment.  It thus extends the observer-relativity of quantum states common to Everett's ``relative state'' interpretation \cite{everett:57}, Rovelli's ``relational'' interpretation \cite{rovelli:96} and Fuch's quantum Bayesianism \cite{fuchs:10} to an observer-relativity of quantum \textit{systems} ({cf.} Zanardi \cite{zanardi:01} where observer-relative systems are already taken for granted and Dugi\'{c} and Jekni\'{c} - Dugi\'{c} \cite{dugic:06, dugic:08} where they are discussed explicitly).  Moreover, each observer's observational capabilities---in quantum-theoretic language, the observables that each observer is capable of deploying---are exactly specified by the observer-environment interaction Hamiltonian.  Decompositional equivalence entails that this sense of observer dependence is shared by classical and quantum physics.  Many authors from von Uexk\"{u}ll \cite{vonUexkull:57} onwards have emphasized that observation is, by its very nature, observer-dependent in this way.

The sense of observer dependence implied by the Black Box model, and hence the model~itself, is~often criticized as subjectivist or even solipsist.  Fuchs responds to the charge of solipsism with characteristic vigor: ``No agent, no outcome for sure, but that's not solipsism: For, no system, no~outcome either!'' (\cite{fuchs:10}, p. 19, Fig. 5 caption).  What the observer is \textit{capable of} observing is determined by the observer-environment interaction (i.e., the Hamiltonian $H_{OE}$), and in the limit of a very large environment, primarily by the internal dynamics of the observer (i.e., by $H_{O}$).  However, what the observer actually observes at any given instant---the output delivered by the black box in response to any given input---is determined by the environment (i.e., by $H_{E}$).  This is ``subjectivity'' in the sense that an observer can only observe what she is capable of observing, but it is not ``subjectivity'' in the (typically pejorative) sense of an observer that observes only what she expects to observe or, worse yet, only what she wants to observe.  As Moore \cite{moore:56} so clearly emphasized, the essence of the Black Box model is the ever-present possibility of surprise.  

``Subjectivity'' in the sense entailed by the Black Box model is, moreover, consistent with FAPP intersubjectivity and hence FAPP objectivity.  Two very similar observers embedded in a very large world in such a way as to make their respective environments and hence their respective observer-environment interactions very similar can be expected to record very similar observational~outcomes.  If the bandwidths of their respective information channels and hence their effective measurement resolutions are small, their outcomes as recorded may even be identical.  Observer dependence does not mean that agreement about outcomes is impossible.  It rather means, as discussed in more detail in \S 3 and \S 4 below, that such agreement can carry no \textit{ontological} weight.  Nothing the observers can do can guarantee that their observations are caused by the same system; indeed the Black box model entails that their observations are caused by different systems, by their own unique and distinct, albeit very similar, environments.  Each observer has, by definition, her own personal \textit{umwelt} \cite{vonUexkull:57}, and every observation she makes is constrained by it.

\section{No-Go Theorems for Black Boxes}

``No-go'' theorems restrict the inferences that observers can make from either theory or~observations.  If the Black Box model indeed provides a general description of observational interactions with physical systems, it should provide no-go theorems that impose the same restrictions on inferences from theory or observations that the familiar no-go theorems of quantum information theory impose.  This section shows that this is indeed the case.

\subsection{Moore's Theorem (1956)}

Edward Moore proved, in 1956, the fundamental no-go theorem for observers of black boxes (\cite{moore:56}, Theorem 2):

\textbf{Theorem} (Moore, 1956]): Finite observations of a black box are insufficient to determine its machine table.

The ``machine table'' of a black box is the list of all \textit{possible} input-output transitions of the black~box.  The finite list of observed input-output transitions accumulated by an observer of some black box $\mathbf{B}$ is at least a partial machine table of $\mathbf{B}$.  Moore's theorem shows that no such partial machine table can be demonstrated to be complete.  Hence it prohibits the inference of a \textit{complete} machine table from any finite list of observed input-output transitions, and hence from any model based on such a list, no~matter how long this list may be.

\textbf{Proof} (Moore, 1956):
Given any partial machine table $T$ obtained by finite observation of some black box $\mathbf{B}$, it is possible to construct arbitrarily many hypothetical machine tables that contain~$T$.  These~constructed machine tables correspond to black boxes with larger state spaces than the minimal black box $\mathbf{B}_{\mathit{T}}$ required to generate $T$.  Because any of these larger black boxes could have generated $T$, nothing about $T$ is sufficient to determine that the observed black box $\mathbf{B}$ is $\mathbf{B}_{\mathit{T}}$, i.e., nothing about $T$ is sufficient to determine that $T$ is the complete machine table of $\mathbf{B}$.  $\square$

Moore's theorem shows that the very next output from any black box can be a complete surprise.  Any black box could, for example, contain a clock that counts to some large number, after which it enables the execution of a qualitatively different dynamics than that executed previously; alarm~clocks, time bombs, commonplace computer programs and many biological systems are obvious examples.  A~significant special case to which this applies is that of ``tests for contextuality'' such as discussed by Kitto \cite{kitto:14}; no finite sequence of observations during which a system appears to behave in a non-contextual way provides any guarantee that the system will not behave in a contextual way in the future.  One can appeal to Occam's razor and \textit{hypothesize} that the behavior that has already been observed is the only behavior of which a black box is capable, but Moore's theorem reminds us that this is always merely an hypothesis.  The very next observation could prove it wrong.  

Moore's theorem gives hypotheses about the behavior of a black box, whether framed as input-output ``laws'' or as proposed internal mechanisms, exactly the status that Popper \cite{popper:63} gives scientific hypotheses in general: they are candidates for empirical falsification.  They are at best~satisficing, not optimal.  To borrow Bell's \cite{bell:90} somewhat pejorative term, they may be true FAPP, but they can never be considered true \textit{simpliciter}.  They cannot, therefore, support the weight of metaphysical or ontological claims.  As noted above, ontological claims about \textit{physical} boundaries and hence physical structure inside a black box fall immediately afoul of decompositional equivalence.  Moore's theorem generalizes this prohibition of ontology to internal structures described in any~language.  Explicitly hypothetical claims---models---of course suffer no such prohibition.  Provisional or FAPP ``knowledge'' of the workings of a black box---the kind of knowledge represented by not-yet-disconfirmed models, is allowed by Moore as it is by Popper; it is certain or god's-eye knowledge that is prohibited. 

\subsection{A Black Box Cannot Be Bounded}

Moore's theorem provides the basis from which to prove the principle result of this paper: that a black box cannot be bounded.  The informal ``picture'' of a black box found in the classical cybernetics literature (e.g., \cite{ashby:56, moore:56}) or in Figure \ref{fig1} places the observer outside the box.  The box is, in~the canonical~example, a device captured from an enemy and given to an engineer to investigate, or in a design example, a device about which only the input-output interface is known.  In either case, the~box is \textit{bounded}, clearly separate from and causally independent of the other items in the laboratory or workshop.  The following theorem shows that this traditional, intuitively-appealing picture is seriously~misleading.

\textbf{Theorem} (no-boundary): If the observable world contains a black box, it is a black box.

To prove this theorem, it is useful to introduce the following:

\textbf{Definition}: 
Let $\mathbf{B}$ be a black box and $\xi$ be an arbitrarily-chosen observable degree of freedom.  The degree of freedom $\xi$ is \textit{separable} from $\mathbf{B}$ if and only every state of $\mathbf{B}$ is independent of the state of $\xi$.

This is the definition of separability familiar from physics ({cf.} Section \ref{sec2.1}), phrased from the perspective of $\mathbf{B}$.  The proof is now simple: Moore's Theorem shows that knowing that $\xi$ is separable from $\mathbf{B}$ is knowing more than can be known.

\textbf{Proof}: 
If the observable world contains a black box $\mathbf{B}$, but is not itself a black box, it must contain at least one observable degree of freedom $\xi$ that is separable from $\mathbf{B}$.  If this is the case, all states of $\mathbf{B}$ and therefore the complete machine table of $\mathbf{B}$ must be independent of the state of $\xi$.  This, however, is by Moore's theorem information about $\mathbf{B}$ that cannot be obtained by finite observation, contradicting the assumption that $\mathbf{B}$ is a black box.  $\square$

This no-boundary theorem confirms in general---taking it as obvious that the world contains at least one black box---the conclusion reached above from the observer as witness formulation and from decompositional equivalence: each observer obtains information only at her boundary with the surrounding ``environment'',  i.e., the rest of her observable universe.  This environment is a black box.  The systems it contains, if any, are not accessible for direct observation.

The no-boundary theorem appears radical, but it is not.  Indeed it rests not on a strong underlying assumption, but on the absence of one: the proof would not go through if it were permitted to assume, a priori, that systems are separable.  In this case $\mathbf{B}$ could be assumed, a priori, to be separable from its environment and hence from $\xi$.  The traditional picture of a black box as a device resting on an engineer's workbench implicitly makes this assumption; the engineer in this traditional picture does not need to worry about the box being dynamically coupled to his heartbeat or to the phases of the~Moon.  Removing the assumption of separability turns these worries, however far-fetched they may~be, into real possibilities.  If they are to be removed from consideration definitively, not just FAPP, they must be removed by observational evidence.  Moore's Theorem shows that definitive observational evidence of independence cannot be obtained by finite means.

\begin{figure}
\centering
\includegraphics[width=16cm]{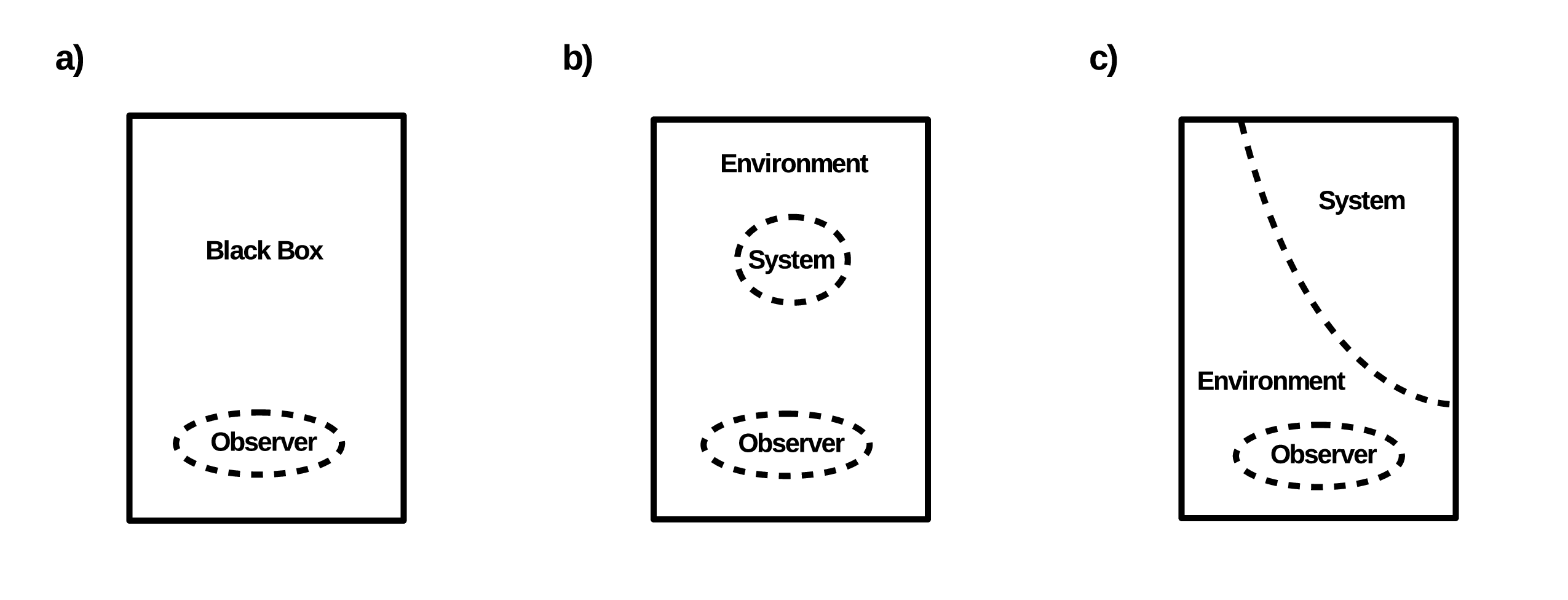}
\caption{The observer within the black box.  (\textbf{a})  The observer is embedded within ``the rest of the observable universe'', which by Theorem 2 is a black box;  (\textbf{b}) The observer is embedded within the environment postulated by the environment as witness formulation of decoherence, from which any information about systems embedded within that environment must be obtained;  (\textbf{c}) The observer is embedded within an observable universe that satisfies decompositional equivalence.  Subsystem boundaries (other than the observer's) within that universe have no effect on the information that the observer can obtain.}
\label{fig2}
\end{figure}

The no-boundary theorem moves each observer into an all-encompassing black box, the rest of that observer's observable universe (Figure \ref{fig2}).  It denies observers independent access to ``systems'' embedded within their observable universes; the physical interaction by which an observer obtains information is defined at the observer---observable universe boundary, and all information obtained by the observer must cross this boundary.  The ``source'' of the information within the box cannot be identified, as doing so would require access to the interior of the box, access that no observer can have.

An immediate consequence of the no-boundary theorem is that the composite observer-box system is, from the observer's point of view, a closed system.  The observer cannot, in particular, detect anything else that interacts with the box; the box is all she sees.  The closure of the composite observer-box system is only a fact FAPP, but it is, for the observer, a conclusion that no observation can~falsify.

The no-boundary theorem does not, of course, limit the observer's ability to \textit{hypothesize} the existence of distinct ``systems'' within the black box that constitutes her observable universe.  Nor~does it imply that such hypotheses cannot be useful FAPP for choosing what inputs to give the box or for predicting what outputs to expect from the box.  It merely establishes the Popperian point that while such hypotheses can be supported by observation, they can never be confirmed by observation.  They~cannot, therefore, support ontological claims.  In particular, they cannot be regarded as ``self-evident truths'' and accorded the status of axioms.  ``Systems exist'' is sometimes regarded as an axiom of quantum theory \cite{zurek:03, fuchs:14}.  As discussed in more detail below, the apparent existence of distinct physical systems should instead be regarded as an explanadum of quantum theory.

\subsection{Corollaries}

The no-boundary theorem shows that separability between any two observable degrees of~freedom, i.e., between any two components of the state(s) of one or more systems, is at best a FAPP assumption.  Separability is, however, the fundamental enabling assumption of the ``classical worldview'' of independently-observable, independently-manipulable physical objects.  Einstein put it this way: ``Without such an assumption of the mutually independent existence (the ``being-thus'') of spatially distant things, an assumption which originates in everyday thought, physical thought in the sense familiar to us would not be possible''.  (quoted in \cite{fuchs:14},  p. 6).  Separability underpins the concept of statistical independence and, therefore, Bell's \cite{bell:64} inequality.  If the observable universe is a black box, the idea that any two observable degrees of freedom would satisfy Bell's inequality is at best a FAPP assumption.  There is no reason for it to be true, and no justifications beyond FAPP utility and convenience for assuming that it is true.  The no-boundary theorem tells us, indeed, to \textit{expect} Bell's inequality to be violated, the surprise elicited in fact by experimental confirmations of its violation by Aspect, Dalibard and Roger \cite{aspect:82} and now many others notwithstanding.  It is important to emphasize that this expectation follows purely and solely from \textit{classical} cybernetics.  No ``quantum'' assumptions are needed.  If ``physical thought in the sense familiar to us'' is not possible without separability, then it is not possible even in \textit{classical} physics.

As shown previously \cite{fields:13a}, the restrictions on inferences from observational outcomes imposed by Bell's theorem \cite{bell:64}, the Kochen-Specker theorem \cite{kochen:67} and the no-cloning theorem \cite{wootters:82} can also be derived from Moore's theorem; they therefore characterize any observations of a black box.  The~no-boundary theorem shows that the restrictions imposed by these theorems characterize observations across the board.  In every case, these restrictions follow from the inability of observers to identify the particular collection of degrees of freedom within the black box under observation---the particular embedded ``system''---that produced a given observational outcome.  If the sources of observational outcomes cannot be singled out, i.e., if separability fails, then outcomes that appear to come from two distinct systems cannot be assumed to be independent (Bell's theorem), outcomes~that appear to come from a single system cannot be assumed to be independent of the order in which they are obtained (Kochen-Specker), and identical outcomes do not indicate the presence of identical systems (no cloning).  This last result is simply a special case of Moore's theorem: no two finite sequences of observational outcomes obtained from two~black boxes can demonstrate that the two~black boxes have the \textit{same} machine table, i.e., no two such sequences of observational outcomes can demonstrate that two~black boxes are clones.  This and the other results are, once again, consequences of classical cybernetics alone.  That they were discovered in the context of quantum systems, and that they are still often regarded as ``intrinsically quantum'' (e.g., \cite{leifer:16} where Bell's theorem is regarded as ``intrinsically quantum'' although the Kochen-Specker and no-cloning theorems are not), indicate the depth with which the view that the observable universe cannot be a black box pervades ``classical'' thinking, i.e., the extent to which the decompositional equivalence of the classical formalism is ignored.

Several additional corollaries follow immediately from the no-boundary theorem.

\textbf{Corollary} (no-communication): 
Two or more observers cannot employ their shared environment as a classical communication channel.

\textbf{Proof}: 
By the no-boundary theorem, the environment shared by two or more observers is a black box.  No observer of this black box can determine the internal source or causal history of any output received from the box.  Therefore, no observer can distinguish states of the box that result from inputs from other observers from states of the box that have no dependence on inputs from the other observers.  $\square$

It cannot, in other words, be determined by finite observation that an output from a black box is a ``message'' from another observer.  Indeed from the point of view of any one observer, all other observers are inside that observer's observable universe and hence inside a black box (Fig. 3). 

\begin{figure}
\centering
\includegraphics[width=13cm]{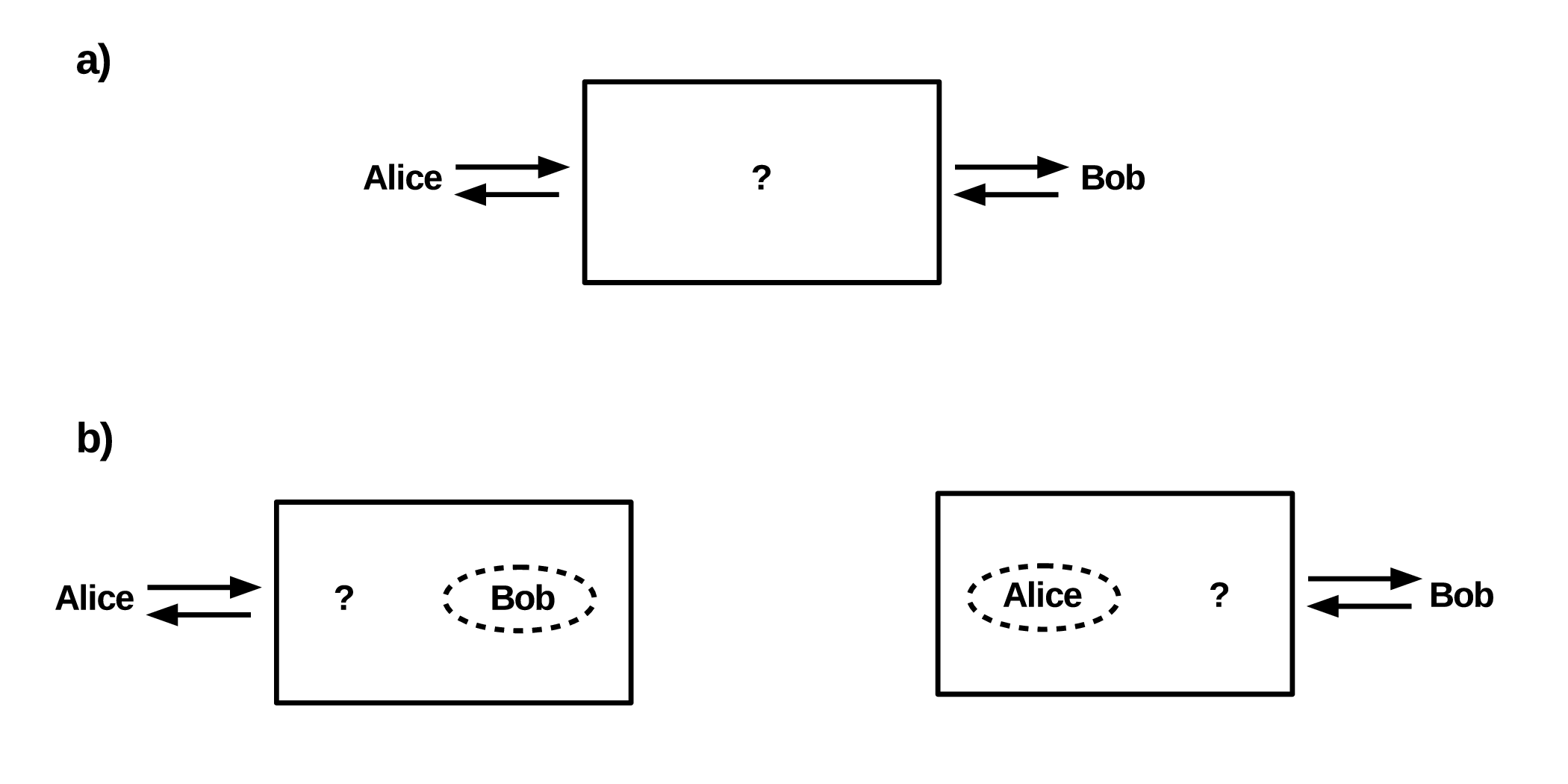}
\caption{(\textbf{a}) Two observers, Alice and Bob, interact with a black box;  (\textbf{b}) It is equivalent to regard Alice as interacting with a black box that contains Bob, and Bob as interacting with a black box that contains~Alice.  It is clear in this latter picture that Alice cannot determine by observation that an output from her box is a message from Bob, nor can Bob determine by observation that an output from his box is a message from Alice.}
\label{fig3}
\end{figure}  

The no-communication corollary is analogous to the no-communication theorem for shared quantum systems \cite{peres:04}, which shows that the local measurement outcomes obtained from a quantum system by one observer cannot depend on the activities of another observer, located somewhere else, who also interacts with the system.  This latter result is generally interpreted as a limitation on \textit{instantaneous} information transfer between observers via the shared quantum system, i.e., as showing that entanglement cannot be used for super-luminal communication.  However, spacelike-separated observers can obtain no direct observational evidence that they are manipulating the same system; it~is for this reason that local observations must be supplemented with classical communication via a separate classical channel in communication protocols (local observations, classical communication or ``LOCC'' protocols \cite{chitambar:14}) that depend on shared quantum systems and hence entanglement as a resource.  If \textit{only} a quantum system is shared, classical communication is impossible and no information can be exchanged, i.e., entanglement ceases to be a resource.  

If observers cannot exchange classical information through a shared black box, they clearly cannot exchange physical systems either.  Physical systems such as clocks, meter sticks and gyroscopes are, however, the \textit{reference frames} with respect to which measurements are made \cite{bartlett:07}.  Hence we have:

\textbf{Corollary} (no-external-reference): 
Observers have no access to external reference frames.

\textbf{Proof}: 
By the no-boundary theorem, any external reference frame is inside a black box and therefore~inaccessible.
$\square$

In particular, observers cannot specifically and exclusively consult external clocks, spatial reference frames, or items of laboratory apparatus, as all of these are internal to the black box with which each observer interacts.  Information from such devices can only be obtained if it is ``packaged'' into an output from the black box.  By the definition of a black box, this packaging process precludes observational access to the source(s) of the packaged information.

The no-external-reference corollary seems radical, but in fact has a simple interpretation: the~reference frames employed by observers to make sense of their observations must be internal.  Without an internal representation of time, for example, the motion of a clock hand has no temporal~meaning; without an internal memory and an internal means of comparing current and remembered events, even the motion itself is undetectable.  Observers, in other words, cannot be ``blank slates'' that just record data if they are to obtain what Roederer has called ``pragmatic information'',  information that enables doing something \cite{roederer:05, roederer:16}.  An observer's internal reference frames determine what kinds of outputs from a black box that observer can observe.  They allow the \textit{interpretation} of observed outputs in terms of hypotheses about the box's states and dynamics.  Without such internal reference frames, the observer cannot do science, indeed the observer cannot \textit{do} anything at all.  It was assumed earlier that observers could choose what to do next; this assumption can now be re-stated as the assumption that all observers of interest have internal reference frames, including in particular an internal time reference frame, that enable them to obtain and act on pragmatic information (cf. \cite{fields:12a}).

If observers have internal structure, it is natural to ask if that internal structure is accessible to other observers.  The no-boundary theorem restricts this question to one case only: the case in which the ``other observer'' in question is the black box with which the observer of interest interacts.  In~this~case, however, the answer has already been decided: the black box obtains only a finite sequence of finite bits strings from the observer.  The situation shown in Figure \ref{fig1}a is, therefore, completely symmetrical: the black box can ``know'' no more about the observer than the observer can know about the black box.  Hence we have:

\textbf{Corollary} (observer-box equivalence): 
Observers are black boxes and vice-versa.

\begin{figure}
\centering
\includegraphics[width=8cm]{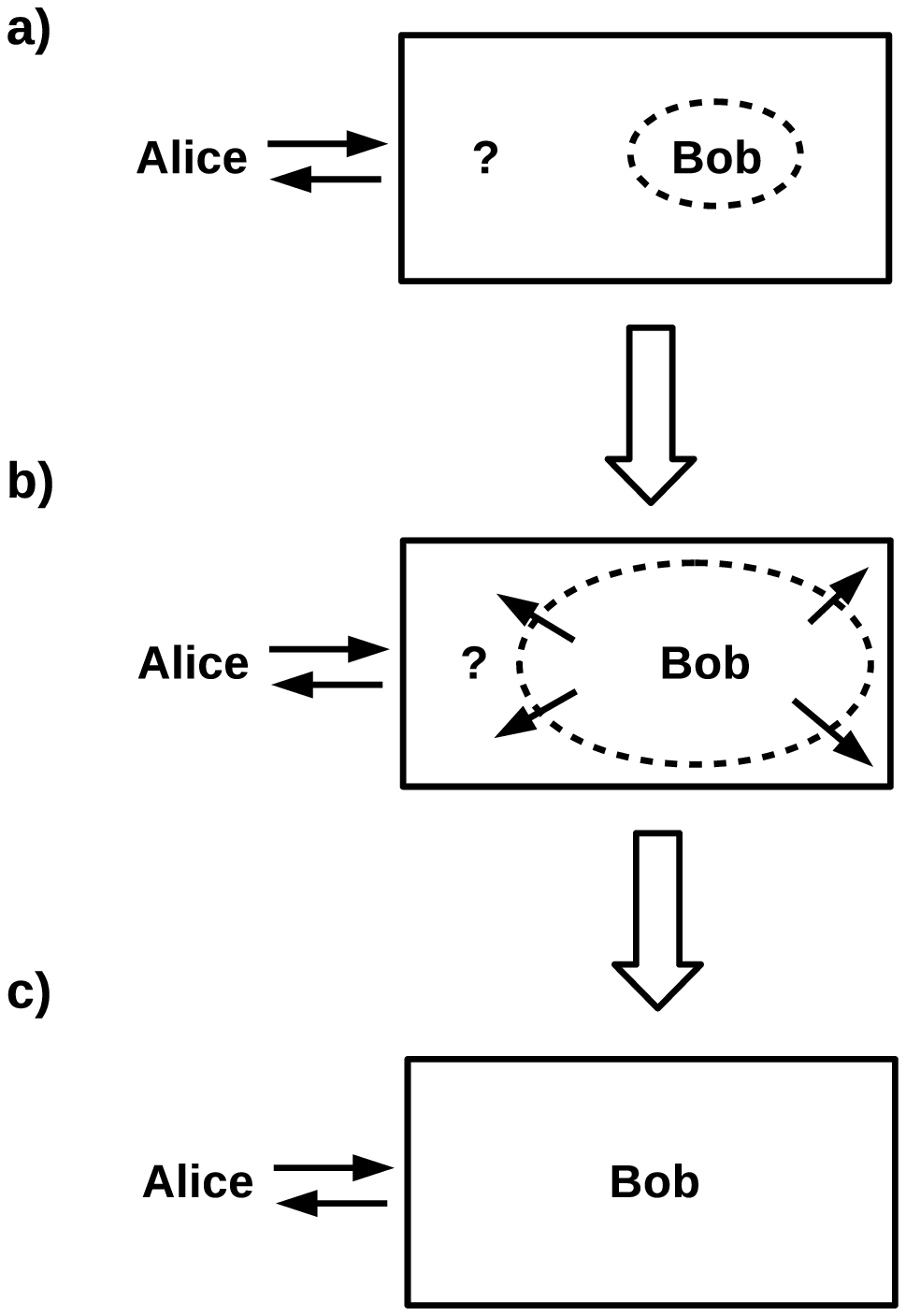}
\caption{(\textbf{a}) Alice interacts with a black box containing Bob, as in Fig. 3b; (\textbf{b}) Bob's boundary within the box can expand arbitrarily without affecting the box's interaction with Alice;  (\textbf{c})  In the limit, Bob~and the box are identical.}
\label{fig4}
\end{figure}

\textbf{Proof}: 
Consider an observer (Alice) interacting with a black box $\mathbf{B}$ that contains another observer (Bob) as shown in Figure \ref{fig3}b.  Because Bob's presence within $\mathbf{B}$ is undetectable by Alice, this~situation is completely general.  By the no-boundary theorem, the outputs from $\mathbf{B}$ cannot depend on Bob's~boundary, therefore it can expand arbitrarily within $\mathbf{B}$ (Figure \ref{fig4}).  In the limit, Bob and $\mathbf{B}$ are identical.  $\square$

This observer-box equivalence corollary extends Ashby's remark about ``real objects or systems'' to include observers, an extension that comports well with Ashby's lifelong commitment to understanding biological intelligence in terms of adaptive information processing \cite{asaro:08}.  It removes, in~particular, any justification for assuming that observers are made out of a different kind of ``stuff'' or have an intrinsically different structure from the environments in which they are embedded.  It allows the boundary around an observer embedded in a black box to be erased; i.e., it recognizes that a black box containing an observer satisfies decompositional equivalence.

\begin{figure}
\centering
\includegraphics[width=10cm]{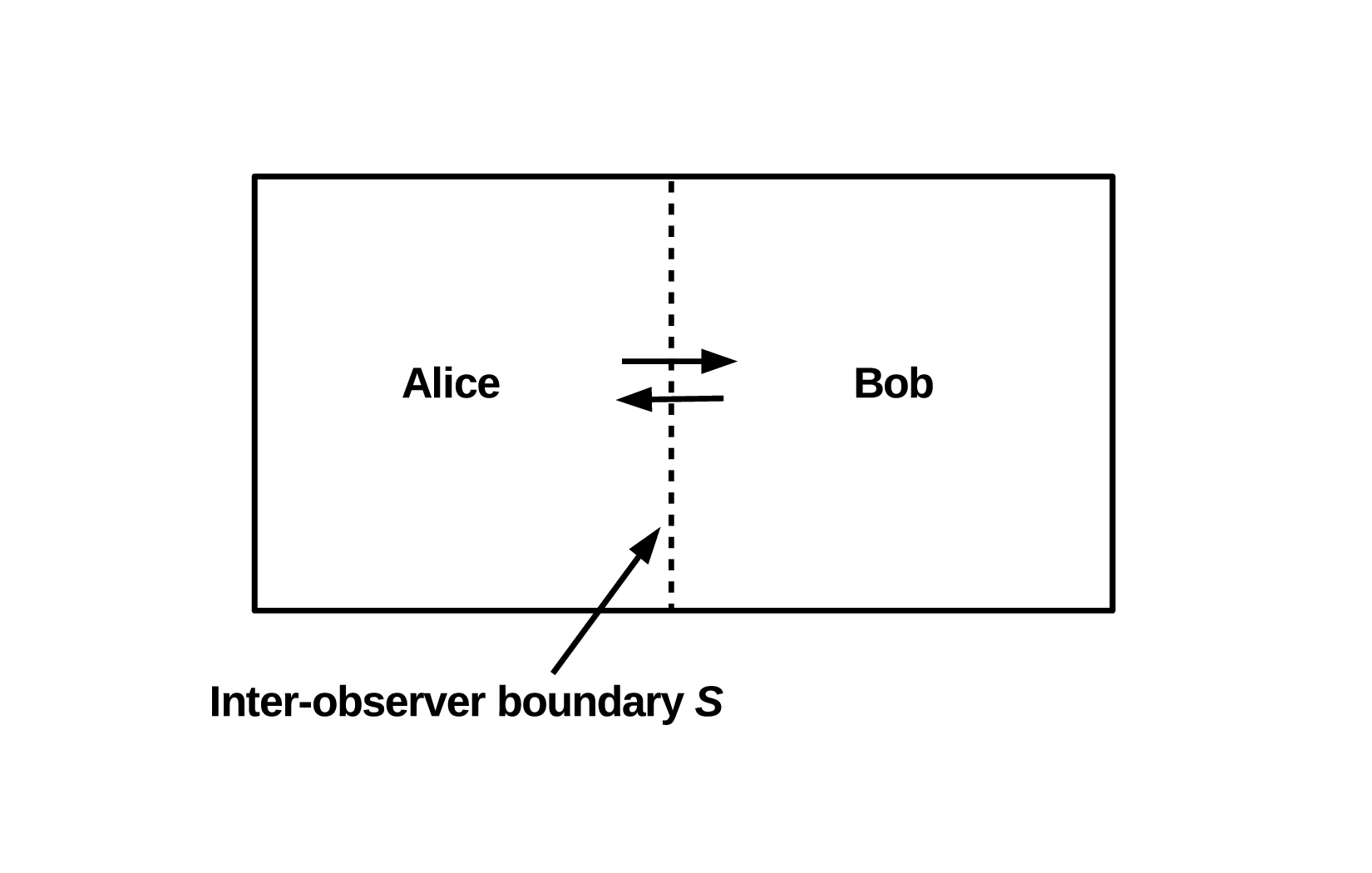}
\caption{Alice and Bob exchange classical information through an inter-observer boundary $S$.}
\label{fig5}
\end{figure}

The observer-box equivalence corollary allows us to re-conceptualize the observer-box interaction as an interaction between two observers or, equivalently, between two black boxes (Figure \ref{fig5}).  The~observers/boxes together comprise a closed system, a ``universe'' in which each observer/box is the other observer/box's ``environment'' or observable universe.  Classical information exchanged by the observers flows through an inter-observer boundary $S$, which plays the role of a classical information channel.  As this boundary is an oriented surface separating the observers, both the information sent and the information received by each observer can be viewed as ``written on'' the side of this surface facing that observer.  Sending and receiving are symmetric, so each observer ``sees'' exactly the same information on their own side of the surface; they differ only in the labelling of what bits have been ``sent'' and what bits have been ``received''.  As there is no process by which to ``unsend'' or ``unreceive'' a message, this ``writing'' on $S$ is permanent.  Indeed the ``memories'' of both Alice and Bob can, without loss of generality, be regarded as implemented by their shared boundary.

If two observers/boxes, e.g., Alice and Bob in Figure \ref{fig5}, are allowed to interact indefinitely, the~amount of information encoded on the inter-observer boundary $S$ will increase, limited only by the assumed-finite energy resources of the observers.  Let $E_{Max} (\mathbf{O})$ be the maximum energy available to an observer $\mathbf{O}$ for information exchange.

\textbf{Definition}: 
Let $\mathbf{A}$ and $\mathbf{B}$ be interacting black boxes that together comprise a closed system, and suppose they are allowed to interact indefinitely.  The \textit{observable machine tables} of $\mathbf{A}$ and $\mathbf{B}$ are then the exactly complementary sequences of inputs and outputs exchanged by $\mathbf{A}$ and $\mathbf{B}$ prior to the smaller of $E_{Max} (\mathbf{A})$ and $E_{Max} (\mathbf{B})$ being~exhausted.

The observable machine table of a black box can clearly be its complete machine table if its number of states is small.  If energy resources are unlimited, the observable machine table of a finite black box is the complete machine table.  If both black boxes are infinite and can access infinite energy, completeness is approached only as a limit.

\textbf{Corollary} (holographic-encoding): 
If a black box is a component of a closed system, its observable machine table can be written on its boundary.

\textbf{Proof}: 
Let $\mathbf{B}$ be the black box in question.  If $\mathbf{B}$ is itself a closed system, it has no boundary but also no inputs or outputs, so it has no observable machine table.  If $\mathbf{B}$ together with one other black box compose a closed system, the result follows by definition of the observable machine table.  If $\mathbf{B}$ interacts with more than one other black box, the no-boundary theorem allows their union to be considered a single black box, so the previous case applies. $\square$

The holographic-encoding corollary provides an alternative way of thinking about the ``inaccessible interior'' of a black box: to say that the interior of a black box is inaccessible is just to say that all available information about the box can be ``read off'' from its exterior.  Physical~systems for which this is true are obviously familiar: they are black holes.  In the case of a black hole, the~information written on the horizon $S$ is proportional to its mass \cite{bekenstein:73}.  In~the case of a black box, the~information written on the horizon $S$ is proportional to its available energy (and hence its mass) or to its behavioral and hence computational complexity, whichever is less.  The connection between~mass, horizon-encoded information and computational complexity has been noted previously in the case of black holes \cite{harlow:13, susskind:16}; it is interesting to see it re-appear here in the purely-classical case of black~boxes.  While in the black-hole case the computational complexity in question is the computational complexity faced by an observer stationed near the horizon who must analyze Hawking radiation from the black hole before it evaporates, in the black box case the computational complexity in question is the computational complexity of the procedure with which the box chooses outputs in response to inputs.  It can be conjectured that the two cases are in fact fully analogous: that the computational complexity of an observer's analysis of Hawking radiation mirrors the computational complexity of  the black hole's ``choice'' of what Hawking radiation to emit.

The holographic-encoding corollary holds for black boxes, or by the observer-box equivalence corollary, observers that are components of closed systems.  As noted earlier, a system being closed can only be a fact FAPP.  All systems \textit{appear}, however, to be closed to observers embedded in them.  Hence~all observers appear, to themselves, to have their observable machine tables encoded as memories on their boundaries.  

It was remarked in the Introduction that modern science assumes that ``there is a way that the world works''. This assumption is naturally interpreted, and historically has been interpreted, as an assumption of determinism.  Standard quantum theory with no \textit{physical} ``collapse'' process is fully unitary and therefore strictly deterministic.  As noted earlier, Tipler has shown that the ``quantum potential'' that the Schr\"{o}dinger equation adds to the classical Hamilton-Jacobi equation is exactly the potential required to make classical physics strictly deterministic \cite{tipler:14}.  Let us assume, therefore, that~the dynamics of any closed system is strictly deterministic.  Even in this case, the following holds:

\textbf{Corollary} (free-will): 
In a closed system comprising two interacting black boxes, neither box determines the other's behavior.

\textbf{Proof}: 
For one box to determine the other's behavior, it must have information specifying the other's state and dynamics.  However, such information cannot be obtained for any black box.  $\square$

This free-will corollary is an analog for black boxes of the Conway-Kochen free-will theorem, which states that the behavior of a quantum system cannot be determined by the information available in its own past lightcone \cite{conway:06}.  While quantum theory can be globally deterministic, no \textit{local} information can be determinative for any system.  It is in this sense of \textit{local} non-determinism that any observer/box can be said to ``freely choose'' its next output (cf. \cite{fields:13b}).  Note that this freedom also applies to each observer's own local information: as each observer's memory contains only its own observable machine table, no observer ``knows'' its own current state or dynamics and hence no observer can locally determine its own next output.  Observers are free and appear autonomous, but they are not autonomous in the literal sense of locally self-determining; local non-determinism includes local non-\textit{self}-determinism.  Local non-determinism prevents strict ``objective'' autonomy for any bounded system, including any observer, but requires apparent autonomy for all observed systems, including  all observers.  Local non-determinism can be stated in terms of decompositional equivalence: any~system's action on its observed world, including any observer's action on her observed world, remains the same when the boundary around the system/observer is erased, rendering the action no longer \textit{the system's/observer's action on the observed world} but rather \textit{the entire world's action on itself}.  Within a black-box based theory, these two actions are indistinguishable.

\subsection{What is Physics about?} \label{sec3.4}

The no-go results demonstrated here, whether in their current classical cybernetic form or in their more traditional quantum-information form, naturally raise the question of what physics is~about.  This~question has traditionally been debated in realist-versus-instrumentalist terms, with the added twist of realists about the ``quantum world'' typically being instrumentalists about the ``classical world'' and vice-versa (e.g., \cite{landsman:07} or for a more recent survey of competing positions, \cite{cabello:15}).  The~no-boundary theorem and the no-communication corollary add a further issue to this debate: that of observations being intrinsically \textit{personal}.  This individuality and principled non-redundancy of observations has previously been emphasized by Fuchs as a central feature of quantum Bayesianism~\cite{fuchs:10}.  If~communications from other observers are regarded as ``observations'' like any others and accorded no special, epistemically-privileged status, the usual notion of objectivity based on inter-subjective agreement breaks down.  It becomes impossible for the agreeing agents to establish unambiguously what they are agreeing about (cf. \cite{fields:14a}).  They can agree FAPP, but they can do no better than this.

The concept of classical information---information that can be irreversibly encoded as a bit string on some physical medium---is taken as a primitive in classical cybernetics and in the current~discussion.  Bits are discrete ``quanta'' of information; hence classical information is itself~``quantized''.  As~noted~above, any observer able to obtain classical information and use it to make predictions must have an internal time reference frame, i.e., an internal clock.  No generality is lost in assuming that the observer's internal clock ``ticks'' whenever an outcome is received from the system being observed; the unit time interval of the observer's internal clock can, therefore, be set to $\Delta t = 1$.  As~the observer must expend a minimum energy $E_{Min} = ln2~kT$ to record each bit, the observer's minimal action per recorded outcome is $E_{Min} \Delta t$.  Hence \textit{the observer's action is quantized} even in classical physics.  Boltzmann's recognition that obtaining information requires expending energy is thus the fundamental idea needed to develop quantum theory.  The quantization $E_{Min} \Delta t ~=~ Constant$ fixes the value of the minimal action; at human physiological temperature, $T = 310$ K, for example, and $\Delta t \sim 200$ ns, the response time of rhodopsin to light at physiological temperature \cite{wang:94}, $E_{Min} \Delta t \sim 6.0 \times 10^{-34}$ J $\cdot$ s, a value remarkably close to Planck's $h \sim 6.6 \times 10^{-34}$ J $\cdot$ s.

As Jennings and Leifer have shown \cite{leifer:16}, the quantization of the observer's action is sufficient to understand the non-commutativity of position and momentum measurements and the position-momentum uncertainty principle, as well as no-cloning, Kochen-Specker contextuality, and~other ``quantum'' effects.   While Jennings and Leifer do not ask what additional assumptions are needed to obtain all of quantum theory within a classical setting, others have done so.  As~Tipler~shows, requiring time evolution to be strictly deterministic is sufficient \cite{tipler:14}; Bohmian mechanics provides a similar demonstration from a different starting point, that of preserving quantum behavior while assuming an ontology of classical ``particles'' \cite{bohm:87}.  Motivated in part by Wheeler's provocative ``it from bit'' proposal \cite{wheeler:90}, a number of information-theoretic \cite{clifton:03, ariano:11, chiribella:11, chiribella:12, hardy:15, muller:15} or more~generally, operational~\cite{knuth:14, deutsch:15, hohn:15} axiomatizations of quantum theory have been proposed.  Within~these formulations, ``physical'' interactions become transformations of bits, typically with qubits as ``inside the box'' intermediaries.  Hence physics in these formulations is ``about'' computation, typically quantum computation.  Reviewing ``device-independent'' operational methods originating in quantum cryptography, which~disallow assumptions about the degrees of freedom or dynamics of the systems from which observational outcomes are obtained, Grinbaum goes even farther, concluding that physics is ``about languages'' (\cite{grinbaum:15}, p. 14), specifically, languages as formal entities without imposed~semantics.

Physics being about information, computation or languages is a far cry from the more traditional, and far more common, view that physics is about the everyday objects of ordinary human experience, even if its theories tend toward the abstruse and postulate non-ordinary objects that are not directly~observed.  \textit{Effectively} embedding the human observer into the theory requires explaining, within the theory, why the everyday observations made by human observers can be described so readily in terms of bounded, causally-independent objects.  The next section makes a step in this direction by showing that ``objects'' can be considered to be collections of outcome values that are held fixed while making measurements.  ``Objects'' are thus internal reference frames.  It is then shown in Section \ref{sec5} that objects construed in this way are the natural external referents of the ``object tokens'' that have been proposed as components of human episodic memories \cite{zimmer:10, fields:12c}.

\section{``Objects'' within a Black Box}

The world may be a black box, but when human observers look at it, they see discrete objects.  The ``environment'' in which these objects are embedded is effectively invisible, a condition that has~obtained, cosmologically, since photons decoupled from matter \cite{harrison:73}.  Hence while human observers physically interact with the environment, predominately the ambient photon field, they~\textit{see}~objects.  From the present perspective, the question of interest is how to describe these objects in a way that is consistent with the Black Box model.

\subsection{Object Identification in Practice}

The almost-automatic human assumption that distinct objects exist in an observer-independent ``objective'' way within the world diverts attention away from a question that is critical for understanding observation: how does an observer \textit{identify} the object being observed?  How, for~example, does one identify a familiar object like a coffee cup?  Such objects do not appear to us in isolation; they must be picked out from scenes that contain many other objects.  The only incoming information available for doing so is that contained in the observational outcomes obtained from the scene as a~whole.

As will be discussed further in Section \ref{sec5} below, answering this system-identification question is non-trivial \cite{fields:16b}.  For the present purposes, however, the most important part of the answer is that observers must dedicate some of the observational outcomes that they associate with an object to its~identification, leaving the rest to indicate its state.  A coffee cup, for example, may be identified by its fixed size, shape and color, while its location and whether it is filled with coffee are allowed to vary, i.e., to be state variables.  What is critical is that \textit{not all observables associated with an object can simultaneously be state variables}; some must be set aside for object identification.  Coffee cups are in this sense like electrons: some observable properties must be regarded as definitional and hence invariant in order for system identification to succeed.

Recognizing that system identification requires a designated collection of observables that are fixed and invariant destroys a key component of the ``classical worldview''.  Ollivier, Poulin and Zurek, for example, operationally define \textit{objectivity} thus (\cite{zurek:05}, p. 3):

\begin{quote}
``A property of a physical system is \textit{objective} when it is:
\begin{enumerate}
\item
simultaneously accessible to many observers,
\item
who are able to find out what it is without prior knowledge about the system of interest, and 
\item
who can arrive at a consensus about it without prior agreement.''
\end{enumerate}
\end{quote}

This definition clearly applies to no systems at all, as observers ``without prior knowledge about the system'' would have no designated observables with which to identify it and observers ``without prior agreement'' about a designated collection of identifying observables would have no basis for any subsequent agreement.  The no-external-reference corollary shows that the reference frames required to characterize system states must be included within the ``prior knowledge'' of observers; inter-observer agreement about reference frames must, similarly, be included among the ``prior agreements'' between~observers.  Shared conceptual schemes and shared category hierarchies are examples of such prior agreements.  As discussed in more detail below, the collection of observable properties that are regarded as defining an object can be considered a reference frame for characterizing that object: without agreement about which observables constitute a reference frame for identifying an~object, multiple observers cannot agree unambiguously about object identification.  As noted earlier, a critically important special case is the collection of reference frames that observers use to identify each other, including their prior agreements about how they will communicate.

\subsection{Superpositions Encode the Unresolvable Ambiguity of Object Identification}

\begin{figure}
\centering
\includegraphics[width=8cm]{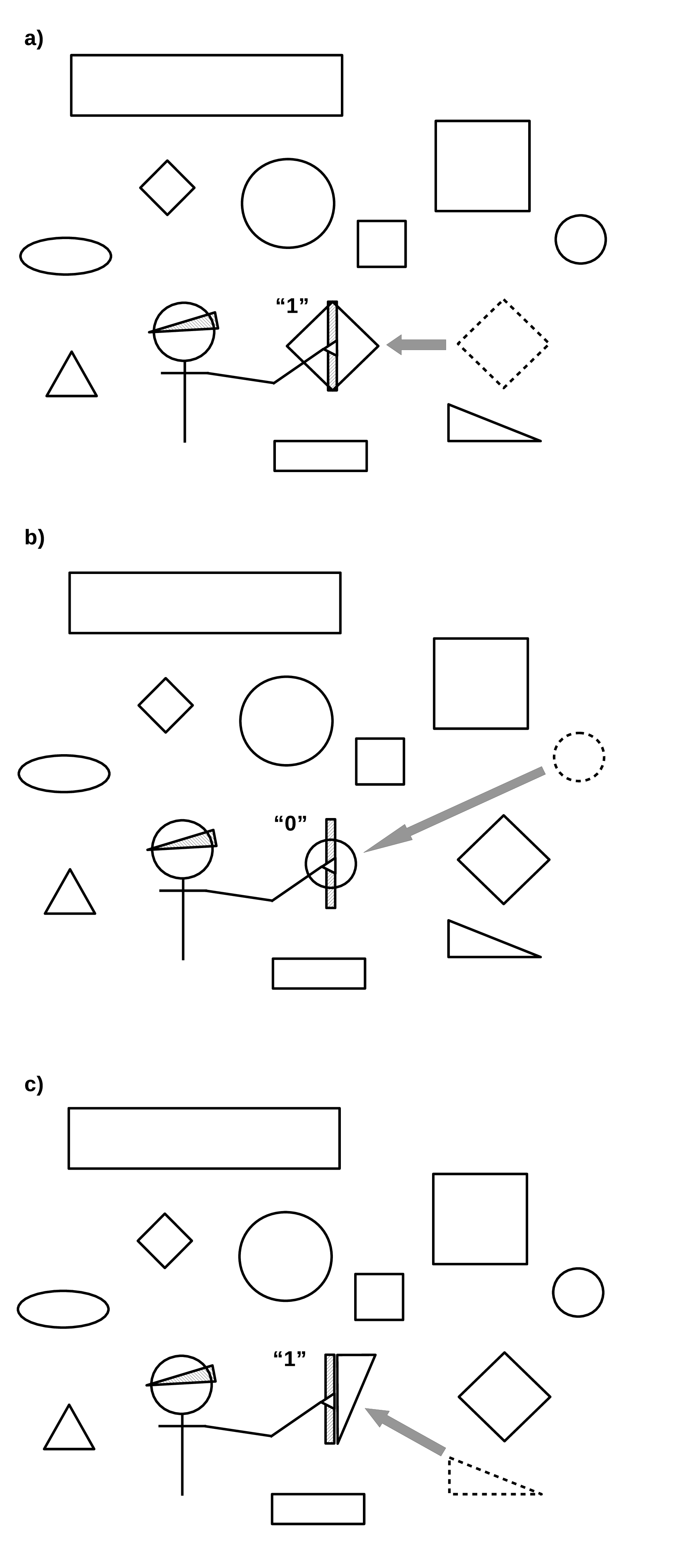}
\caption{(\textbf{a}) An observer equipped with a meter stick and embedded in a world of multiple objects.  The meter stick reports an outcome of ``1'' whenever an object with a linear dimension of 1 m is placed in contact with it.  The observer has no ability to observe any features of the objects other than the linear dimension reported by the meter stick;  (\textbf{b}) The meter stick reports an outcome of ``0'' whenever an object with a linear dimension other than 1 m is placed in contact with it;  (\textbf{c})  A mechanism that places successive objects in contact with the meter stick produces a stream of binary outcome values.}
\label{fig6}
\end{figure}

The role of observational outcomes in identifying objects can be illustrated with a simple thought~experiment.  Consider an observer equipped with a single reference frame, a vertically-oriented meter stick, who is embedded in a world containing multiple objects, some of which have some linear dimension equal, within the resolution of the meter stick, to 1 m while others do not (Figure \ref{fig6}).  The meter stick registers a ``1'' outcome if it comes into contact with an object having a vertical dimension of 1 m and registers a ``0'' outcome if it comes into contact with an object having a different vertical dimension; it registers no outcome if it is not in contact with an object.  An unseen mechanism occasionally puts one of the objects in contact with the meter stick.  The observer receives no information other than the ``1'' and ``0'' outcomes from the meter stick; in particular, the observer is unable to observe the size, shape, orientation, or anything else about the objects other than the vertical linear dimension reported by the meter stick.  The world of this observer is clearly a black box.

With what system does the observer interact when an outcome value of ``1'' is received?  The~\textit{observed state} of that system is known: it is whatever state yields an outcome value of ``1.''  Call~this state ``$|1\rangle$'' and call its complement, the state that yields an outcome value of ``0'', ``$|0\rangle$.''  What, however, is the system that occupies these states?  While the no-boundary theorem gives the precise answer that the ``system'' is the black-box world as a whole, the human tendency to associate distinct observational outcomes with distinct systems suggests an answer postulating two systems, an ``object'' $\mathbf{S}_{\mathrm{1}}$ perpetually~in, and hence identified by $|1\rangle$ and a second ``object'' $\mathbf{S}_{\mathrm{2}}$ perpetually in and identified by $|0\rangle$.  These two alternatives are operationally equivalent: interacting with a single system that can occupy two states is indistinguishable from interacting with two systems, each of which permanently occupies a single state. 

From the ``god's eye'' perspective of Figure 6, one can associate multiple distinct objects distinguished by the unobserved and hence ``hidden'' variable of geometric shape with both $\mathbf{S}_{\mathrm{1}}$ and~$\mathbf{S}_{\mathrm{2}}$.  Providing the observer in Figure 6 with a shape detector would allow a choice between identifying objects by size and treating shape as a state variable or identifying objects by shape and treating size as a state variable.  These descriptions are, once again, operationally equivalent, and both are equivalent to a description in which there is only one object---the black box---and both size and shape are state variables.  The only basis for predictions in any of these descriptions is the sequence of bit strings observed so far, a sequence that provides no information beyond a lower bound on complexity about the mechanism that selects objects to put in contact with the meter stick.  In particular, the observations provide no basis for deciding whether the mechanism selects ``objects'' for presentation at random or by following a deterministic algorithm.

It is natural for an observer who knows no more about what is being observed than the observational outcomes obtained thus far to represent the source of these outcomes, whatever it~is, by a linear superposition of the observed outcomes.  As distinct outcomes can indicate either distinct states or distinct systems, this representation can be interpreted as either a superposition of states or as a superposition of systems; the two interpretations are entirely equivalent \cite{fields:16a}.  While the former is by far the more familiar, the latter is implicit in Feynman diagrams and is even more evident in more abstract representations, such as the amplituhedron \cite{arkani:14}, in which particles and even spacetime are elided altogether.

\subsection{Objects as Internal Reference Frames}

It is increasingly recognized that the reference frames employed to make physical measurements are not mere abstracta but rather are physical objects \cite{bartlett:07}.  Successfully employing a LOCC protocol with entangled photons, for example, requires sharing not just a \textit{description} of the measurement conditions at the two spatially-separated sites, but also a physical orientation reference frame such as the Earth's gravitational field.  The physicality of reference frames forces the physicality of instrument calibration procedures to be taken explicitly into account; when this is done, the standard ``quantum'' uncertainty relations emerge even for purely classical measurements \cite{krechmer:16}.  The complementarity between system-identifying observables and state observables makes it clear why this must be the case across the board: one or the other set of observables must be measured first, and the finite time required for measurement is time during which the unmeasured observables are free to undergo unpredictable dynamic evolution.

The no-external-reference corollary shows, however, that reference frames cannot be \textit{external} physical objects; they must be internal to the observer.  Even an external calibration standard requires identification by observation: from the observer's perspective, it is merely a collection of observational~outcomes.  Calibration itself is a process of comparing observational outcomes that identify and characterize the state of the ``standard'' to observational outcomes that identify and characterize the state of the ``instrument'' being calibrated.  This comparison is an operation on bit strings carried out by the observer; it is a computation.  It is essential to the meaning of this computation that the outcomes that identify both the calibration standard and the instrument remain invariant during the calibration process.  As with the fixed size, shape and color of a coffee cup that enable reaching for it, grasping it, and filling it with coffee, it is the invariance of outcomes that identify objects with respect to outcomes that indicate changes of state that allow state changes to be recognized as~such.

If all objects, including objects that function as calibration standards and reference frames, are~identified by sets of observational outcomes that remain invariant while other relevant observational outcomes change, it is natural to consider these sets of invariant outcomes to be the \textit{internal reference frames} that the no-external-reference corollary requires.  These internal reference frames are information-bearing structures---effectively, data structures---that encode the observer's expectations about the objects present ``within'' the black box under observation.  In~Bayesian predictive-coding \cite{friston:10} terms, they are prior probability distributions.  Without such priors, the~observer cannot identify objects, and hence cannot recognize state changes in objects.

There is, however, no need to postulate the real existence of external objects at all; the~no-boundary theorem guarantees that their existence cannot be revealed by observation.  Any observer's observational outcomes are generated by that observer's impenetrable black box.  The observer encoding a set of internal reference frames that identify objects is all that is required to explain the \textit{appearance}, to the observer, of objects within the box (von Foerster \cite{vonFoerster:76} refers to such internal reference frames as ``eigenforms''; see also \cite{kauffman:03, kauffman:05}).  Hence the observer encoding a set of internal reference frames is all that physics needs postulate.  Internal reference frames are necessary for object perception in a black-box world; they are also sufficient.

\section{The Black Box as a Cross-Disciplinary Paradigm}\label{sec5}

The assumption that the sources, within the world, of observational outcomes can be spatially localized and bounded is central to the classical worldview.  It goes hand-in-hand with the assumption that specific states of spatially localized, bounded systems can be ``prepared'' by local manipulations.  The classical cybernetics of the mid-20th century was by no means the first or the only intellectual movement to question these assumptions.  Philosophers have criticized them at least since Heraclitus.  The 20th century saw, however, an explosion of criticism of these assumptions that ranged across scientific disciplines.  As science requires replicable experiments performed on shareable apparatus and reported on shareable media, this burst of criticism of the assumptions on which these conditions rest indicate that the logic of science itself is at most paraconsistent \cite{dietrich:15}.

\subsection{Formal Semantics, Device Independence and Virtual Machines}

By the 1930s, Wittgenstein's essentially 19th-century conception of propositions in natural or formal languages ``picturing'' unique relations between independently-existing objects \cite{witt:22} had been largely replaced, in no small part due to Wittgenstein's own efforts, by a radically different conception of languages and of symbol systems generally as supporting arbitrarily many distinct interpretations.  This view reached maturity in Tarski's development of formal model theory as a general approach to semantics \cite{tarski:44}; however, it is evident in the Turing machine \cite{turing:36} and in the work of G\"{o}del, Church~and others.  Via the later work of Quine and others, the disconnection of semantics from syntax imposed by formal model theory informs all recent thinking about natural languages.  The implications of this break for ontology have not gone unnoticed.  Quine, for example, points out that with this disconnection, ``physical objects are postulated entities which round out, and simplify our account of the flux of experience ... the conceptual scheme of physical objects is a convenient myth, simpler than the literal truth'' (\cite{quine:48}, p. 32).  Physical objects are, in other words, for Quine just \textit{semantic interpretations} of the observational outcomes that constitute ``the flux of experience.''  Experience itself is experience of a black box.

The most powerful practical impact of the idea that symbols could support arbitrarily many distinct interpretations was in computer science; indeed it can be argued that the disconnection of syntax and hence implementation from semantics is the foundational idea of computer science.  It~enables the two defining characteristics of post-1950s computing: device independence (the same software can run on arbitrarily many distinct hardware platforms) and virtual machines (a single hardware system, appropriately programmed, can exhibit the input-output behaviors of arbitrarily many distinct devices).  These two concepts are closely coupled: virtual machines, in the form of interpreters and compilers, enable device independence, while device independence is what makes virtual machines so useful and valuable \cite{tan:76, smith:05}.  Without these concepts, we would still be programming special-purpose computers by re-arranging their hardware.   With them, we are able to regard a device that fits comfortably in a shirt pocket as performing the specialized functions of hundreds to thousands of different tools.

Despite its everyday familiarity within computer science, the virtual machine concept has yet to penetrate the disciplinary discourse of physics.  ``Device-independent methods,'' however, are virtual machine methods.  An experimenter who ``prepares'' the state of a quantum system is in fact preparing the state of a virtual machine, often literally in a computer-controlled laboratory.  ``Experimental data''~are, similarly, outputs of virtual machines, again often literally.  From the experimenter's point of view, these virtual machines are black boxes, in many cases black boxes that are literally functionally interchangeable, and in practice regularly interchanged, with other black boxes that offer the same functionality FAPP.  What goes on in the interiors of these black boxes is unknown, and in many cases arguably unknowable, even in principle, by human beings \cite{partridge:10}.  To do an experiment is, therefore, to undertake an extended exercise in \textit{semantic interpretation} of observational outcomes obtained from unknown and (at least FAPP but very possibly in principle) unknowable sources.  While some ``physical reality'' or other must be postulated for the exercise to make sense, the observational outcomes obtained leave this reality arbitrarily underspecified.  All of physics, in other words, is device-independent by its very nature.  Experimentally-accessible ``physical systems'' are bits of semantics, or as Grinbaum~\cite{grinbaum:15} puts it, bits of language.

\subsection{Cognitive Science and AI}

The 1960s ``cognitive revolution'' in psychology reintroduced the idea that understanding what is going on inside the subject's---i.e., the observer's---head is important for understanding the subject's~behavior.  Mere stimulus-response laws are not sufficient \cite{chomsky:59}.  Cognitive subjects are, however, notoriously black boxes: the observational tools available to experimental psychology and neuroscience do not directly reveal either the contents of cognition or the processes that manipulate those contents.  In particular, they do not reveal the relationship between cognitive states and states of the external world.  The question of whether this semantic relationship can be considered fixed by causal processes and hence unique underlies the ``symbol grounding problem'' (SGP) \cite{harnad:90, taddeo:05} in cognitive science and AI.  As ``grounding'' a symbol in the external world requires identifying the object, event, or class of objects or events to which it refers, the \textit{external} SGP is equivalent to the problem of object or system identification \cite{fields:14b}.  Responses to the SGP range from treating the observer's world as a black box, at least methodologically \cite{fodor:80}, to considering essentially arbitrary components of the state of the world as components of the observer's cognitive state \cite{clark:98}.  The former response replicates the conceptual structure of Figure \ref{fig5}; the latter reinforces the arbitrariness with which the observer-world boundary is drawn.

While symbol grounding can be viewed as a purely theoretical or even philosophical problem in cognitive science, it becomes a practical problem in AI, particularly for ``embodied'' and ``situated'' systems such as mobile robots \cite{anderson:03}.  In ``open'' task environments in which not all objects can be identified using a priori knowledge, object identification criteria and hence the ``grounds'' for object-specifying symbols must be learned.  How such criteria can be learned, even~just~FAPP, is a central problem in both developmental robotics \cite{cangelosi:15} and developmental psychology \cite{fields:16b}.  \textit{Internally}~grounding object symbols in object-appropriate, context-relative perceptual-motor associations provides a FAPP solution to the SGP; this solution effectively replicates the embedding relationship between object tokens \cite{zimmer:10} and event files \cite{hommel:04} or episodic memories~\cite{eichenbaum:07} in theories of human object recognition \cite{fields:12c}.  The external or ontological version of the SGP is widely regarded as both unsolvable and irrelevant to practical robotics \cite{cubek:15}.  Rejecting the external SGP as a problem for robotics is considering the world in which a robot behaves to be a black box.

\subsection{Evolution and Development}

Cognitive scientists who reject a black-box view of the observer-world interaction often appeal to biological evolution as a mechanism for enforcing a causal connection between world states and cognitive states.  Mark, Marion and Hoffman have, however, shown using evolutionary game simulations that agents responsive only to fitness payoffs out-compete agents responsive to the actual distribution of resources in a simulated world \cite{mark:10}.  This result motivates the ``interface theory of perception'' (ITP), which conceptualizes perceptual systems as providing finite-resolution information about the fitness consequences of actions in a black-box world \cite{hoffman:15}.  Within ITP, the~semantic relation between the agent's internal representations and the world states or structures to which fitness consequences are associated is essentially arbitrary, as expected within a model-theoretic or virtual-machine formulation of the semantics.  The principle of ``conscious realism'' of Hoffman and Prakash \cite{hoffman:14}, which treats an agent's world as another agent and hence treats agent-world games as two-agent games, replicates the symmetric structure of Figure \ref{fig5} within ITP.

While ITP is explicit in conceptualizing perception and hence cognition as interaction with a black box, it is not alone in making this assumption.  The Bayesian predictive coding framework~\cite{friston:10} provides an organizing principle for not only neurocognitive architecture \cite{friston:11}, but for the evolution and development of complex biological structures in general \cite{friston:15}.  As Friston {et al.} point out, the~predictive coding framework ``only make(s) one assumption; namely, that the world can be described as a random dynamical system'' (\cite{friston:15}, p. 9).  The world is, in other words, from the observer's perspective a black box.  The boundary of the box in this formalism is the ``Markov blanket'' of the observer; the observer effectively perceives and acts on only the inside surface of this blanket~\cite{friston:13}, making it an ``interface'' in the sense employed by ITP.  The interactions of the observer with the blanket are highly dependent on the observer's Bayesian priors, i.e., on the observer's conceptual and categorical structures; hence they display significant contextuality of the second kind identified by Kitto \cite{kitto:14}.  Theoretical approaches to conditions such as autism within the predictive coding framework (e.g., \cite{lawson:14}) emphasize this contextuality.

\section{Conclusions}

As shown here, the over 60-year-old concept of the Black Box provides a powerful basis for conceptualizing observer-system interactions.  While traditional discussions of the Black Box placed the observer outside the box, the no-boundary theorem moves the observer inside the~box.  Moving~the observer inside the box, and then recognizing that the observer-box boundary is itself arbitrarily~movable, leads immediately to no-go theorems of the type familiar from quantum information theory.  The observer-box equivalence corollary then shows that observers are themselves black boxes, while the holographic-encoding corollary shows that both observers and their observed worlds are analogous, from an information-theoretic perspective, to black holes.  Physics, therefore, already has a detailed and well established theory of the observer: the theory of the black hole.  Conceptualizing the observer as a black hole forces observer-observer interactions to be regarded as physical interactions occurring at a defined interface between otherwise mutually-inaccessible systems.  It forces us to acknowledge that we do not know what Schr\"{o}dinger saw when he opened his infamous box and looked for his cat.  We only know what he \textit{reported} seeing.  Indeed we each only know \textit{our own interpretation}, based on our own physical interaction with his physically-tokened report, of what he reported seeing.  As Fuchs has pointed out, with this level of epistemic personalization the usual ``measurement problem'' of quantum theory simply disappears \cite{fuchs:10}.  It is no longer unclear why observers report classical outcomes: only classical outcomes---outcomes encodable by classical bit strings---are recordable in a thermodynamically-irreversible way, and hence only classical outcomes are reportable.  It is no longer unclear how the measurement basis is chosen: observers deploy the observables they are able to deploy, represented in the measurement bases they are able to deploy.  The traditional measurement problem is replaced, however, by an even deeper problem: a~deep and in-principle unresolvable ambiguity about both the ontological structure and the state of the world.  This deep ambiguity is not a ``quantum'' phenomenon.  It follows solely from classical physics, the~classical physics of the Black Box.  Had Moore's theorem been proved 20 years before the development of quantum theory instead of 20 years after, this deeper problem may have been considered the ``measurement problem'' all along.

Taking the Black Box seriously as a model of observation not only clarifies the relation between classical and quantum physics; it also clarifies the relation between physics and computer science, cognitive science and even biology.  It forces physics to view semantics in model-theoretic terms, and~thereby enables physics to view physical systems in virtual-machine terms.  While this view is implicit in computational conceptualizations of physical dynamics, it typically only appears explicitly in speculations about simulated worlds (e.g., \cite{bostom:03}).  The simulation assumption is, however, not~required for the conclusion that any observer's observed world is a virtual world.  This conclusion requires only that the world itself, the world in which the observer is embedded, is a black box.
\vspace{12pt}

\section*{Acknowledgement} 

This work was supported in part by The Federico and Elvia Faggin Foundation.  Conversations with Eric Dietrich, Federico Faggin, Don Hoffman, Ken Krechmer, Mike Levin, Chetan Prakash and Manish~Singh, as well as the comments and criticisms of two anonymous referees have helped to clarify the ideas in this paper.


\begin{thebibliography}{999}

\bibitem{ashby:56}
Ashby, W.R.  {\em Introduction to Cybernetics};  Chapman and Hall:  London, UK, 1956.

\bibitem{landauer:61}
Landauer, R.  Irreversibility and heat generation in the computing process. {\em IBM J. Res. Dev.} {\bf 1961}, {\em 5}, 183--195.

\bibitem{landauer:99}
Landauer, R.  Information is a physical entity. {\em Physica A} {\bf 1999}, {\em 263}, 63--67.

\bibitem{bennett:03}
Bennett, C.H.  Notes on Landauer's Principle, reversible computation, and Maxwell's Demon.  {\em Stud. Hist. Philos. Mod. Phys.} {\bf 2003}, {\em 34}, 501--510.

\bibitem{baccia:06}
Bacciagalupi, G.; Valenti, A.  {\em Quantum Theory at the Crossroads: Reconsidering the 1927 Solvay Conference};  Cambridge University Press: Cambridge, UK, 2006.

\bibitem{landsman:07}
Landsman, N.P.  Between classical and quantum.  In {\em Handbook of the Philosophy of Science: Philosophy of Physics}; Butterfield, J., Earman, J., Eds.; Elsevier: Amsterdam, The Netherlands, 2007;  pp. 417--553.

\bibitem{wallace:08}
Wallace, D.  Philosophy of quantum mechanics. In {\em The Ashgate Companion to
Contemporary Philosophy of~Physics}; Rickles, D., Ed.;   Ashgate: Aldershot, UK, 2008;  pp. 16--98.

\bibitem{norson:13}
Norsen, T.; Nelson, S.  Yet another snapshot of foundational attitudes toward quantum mechanics.
2013, preprint arXiv:1306.4646v2.

\bibitem{schloss:13}
Schlosshauer, M.; Kofler, J.; Zeilinger, A.  A snapshot of foundational attitudes toward quantum
mechanics. {\em Stud. Hist. Philos. Mod. Phys.} {\bf 2013}, {\em 44}, 222--223.

\bibitem{sommer:13}
Sommer, C.  Another survey of foundational attitudes towards quantum mechanics. 2013, preprint arXiv:1303.2719v1.

\bibitem{fuchs:10}
Fuchs, C.  QBism, the perimeter of Quantum Bayesianism.  2010,  preprint arxiv:1003.5201v1.

\bibitem{cabello:15}
Cabello, A.  Interpretations of quantum theory: A map of madness. 2015, preprint arXiv:1509.04711v1.

\bibitem{bell:90}
Bell, J.S.  Against measurement.  {\em Phys. World} {\bf 1990}, {\em 3}, 33--41.

\bibitem{fuchs:16}
Fuchs, C.  On participatory realism.  2016, preprint arxiv:1601.04360v1.

\bibitem{wheeler:83}
Wheeler, J.A.  Law without law.  In  {\em Quantum Theory and Measurement};  Wheeler, J.A.,  Zurek, W.H., Eds.; Princeton University Press: Princeton, NJ, USA, 1983;  pp. 182--213.

\bibitem{vonUexkull:57}
Von Uexk\"{u}ll, J.  A stroll through the worlds of animals and men. In {\em Instinctive Behavior}; Schiller, C., Ed.; van~Nostrand Reinhold: New York, NY, USA, 1957;  pp. 5--80.

\bibitem{vonFoerster:76}
Von Foerster, H.  Objects: Tokens for (eigen-) behaviors.  {\em ASC Cybern. Forum} {\bf 1976}, {\em 8},  91--96.

\bibitem{kampis:96}
Kampis, G.  Explicit epistemology.  {\em Revue de la Pensee d’Aujourd’hui} {\bf 1996}, {\em 24}, 264--275. 

\bibitem{koenderink:14}
Koenderink, J.  The all-seeing eye.  {\em Perception} {\bf 2014}, {\em 43}, 1--6.

\bibitem{rossler:87}
R\"{o}ssler, O.E. Endophysics. In  {\em Real Brains---Artificial Minds}; Casti, J., Karlquist, A., Eds.;  North-Holland:  New~York,  NY, USA, 1987;  pp. 25--46.

\bibitem{vonNeumann:55}
Von Neumann, J.  {\em The Mathematical Foundations of Quantum Mechanics};   Princeton University Press:  Princeton, NJ, USA, 1955.

\bibitem{pattee:01}
Pattee, H.H.  The physics of symbols: Bridging the epistemic cut.  {\em Biosystems} {\bf 2001}, {\em 60}, 5--21.

\bibitem{kauffman:15}
Kauffman, S.A,; Gare, A.  Beyond Descartes and Newton: Recovering life and humanity.  {\em Prog. Biophys. Mol.~Biol.} {\bf 2015}, {\em 119}, 219--244.

\bibitem{kitto:14}
Kitto, K.  A contextualised general systems theory. {\em Systems} {\bf 2014}, {\em 2}, 541--565.

\bibitem{tipler:14}
Tipler, F.J.  Quantum nonlocality does not exist.  {\em Proc. Natl. Acad. Sci. USA} {\bf 2014}, {\em 111}, 11281--11286.

\bibitem{bohm:87}
Bohm, D.; Hiley, B.J.; Kaloyerou, P.N.  An ontological basis for the quantum theory.  {\em Phys. Rep.} {\bf 1987}, {\em 144}, 321--375.

\bibitem{rosen:86}
Rosen, R.  On information and complexity.  In   {\em Complexity, Language, and Life: Mathematical Approaches}; Casti,~J.L.,  Karlqvist, A., Eds.;  Springer: Berlin, Germany, 1986;  pp. 174--196.

\bibitem{polanyi:68}
Polanyi, M.  Life's irreducible structure.  {\em Science} {\bf 1968}, {\em 160}, 1308--1312.

\bibitem{shannon:48}
Shannon, C.E.  A mathematical theory of communication. {\em Bell Syst. Tech. J.} {\bf 1948}, {\em 27}, 379--423.

\bibitem{bohr:37}
Bohr, N.  Causality and complementarity.  {\em Philos. Sci.} {\bf 1937}, {\em 4}, 289--298.

\bibitem{landau:77}
Landau, L.D.; Lifshitz, E.M.  {\em Quantum Mechanics: Non-Relativistic Theory};   Pergamon: Oxford, UK, 1977.

\bibitem{mermin:89}
Mermin, D.  What's wrong with this pillow?  {\em Phys. Today}  {\bf 1989},  {\em 42},  9--11.

\bibitem{zeh:70}
Zeh, D. On the interpretation of measurement in quantum theory. {\em Found. Phys.} {\bf 1970}, {\em 1}, 69--76.

\bibitem{zeh:73}
Zeh, D. Toward a quantum theory of observation. {\em Found.  Phys.} {\bf 1973}, {\em 3}, 109--116.

\bibitem{zurek:81}
Zurek, W.H. Pointer basis of the quantum apparatus: Into what mixture does the wave packet collapse? {\em Phys. Rev. D} {\bf 1981}, {\em 24}, 1516--1525.

\bibitem{zurek:82}
Zurek, W.H. Environment-induced superselection rules. {\em Phys. Rev. D} {\bf 1982}, {\em 26}, 1862--1880.

\bibitem{joos-zeh:85}
Joos, E.; Zeh, D. The emergence of classical properties through interaction with the environment. {\em Z. Phys. B Condens.  Matter} {\bf 1985}, {\em 59}, 223--243.

\bibitem{zurek:98}
Zurek, W.H. Decoherence, einselection and the existential interpretation (the
  rough guide). {\em Philos. Trans. R.  Soc. A} {\bf 1998}, {\em 356}, 1793--1821.

\bibitem{zurek:03}
Zurek, W.H. Decoherence, einselection, and the quantum origins of the classical. {\em Rev. Mod. Phys.} {\bf 2003}, {\em 75}, 715--775.

\bibitem{schloss:07}
Schlosshauer, M. {\em Decoherence and the Quantum to Classical Transition};   Springer: Berlin, Germany,  2007.

\bibitem{zurek:04}
Ollivier, H.; Poulin, D.; Zurek, W.H. Objective properties from subjective quantum states: Environment as a~witness. {\em Phys. Rev. Lett.} {\bf 2004}, {\em 93}, 220401.

\bibitem{zurek:05}
Ollivier, H.; Poulin, D.; Zurek, W.H. Environment as a witness: Selective proliferation of information and emergence of objectivity in a quantum universe. {\em Phys. Rev. A} {\bf 2005}, {\em 72}, 042113.

\bibitem{zurek:06}
Blume-Kohout, R.; Zurek, W.H. Quantum Darwinism: Entanglement, branches, and the emergent classicality of redundantly stored quantum information. {\em Phys. Rev. A} {\bf 2006}, {\em 73}, 062310.

\bibitem{zurek:09}
Zurek, W.H. Quantum Darwinism. {\em Nat. Phys.} {\bf 2009}, {\em 5}, 181--188.

\bibitem{zurek:10}
Riedel, C.J.; Zurek, W.H. Quantum Darwinism in an everyday environment: Huge redundancy in scattered~photons. {\em Phys. Rev. Lett.} {\bf 2010}, {\em 105}, 020404.

\bibitem{zurek:12}
Riedel, C.J.; Zurek, W.H.; Zwolak, M. The rise and fall of redundancy in decoherence and quantum~Darwinism. {\em New J. Phys.} {\bf 2012}, {\em 14}, 083010.

\bibitem{zurek:14}
Zwolak, M.; Riedel, C.J.; Zurek, W.H. Amplification, redundancy, and quantum Chernoff information. {\em Phys.~Rev. Lett.} {\bf 2014}, {\em 112}, 140406.

\bibitem{chiribella:06}
Chiribella, G.; D'Ariano, G.M.  Quantum information becomes classical when distributed to many users.  {\em Phys. Rev. Lett.} {\bf 2006}, {\em 97}, 250503.

\bibitem{korbicz:14}
Korbicz, J.K.;Horodecki, P.; Horodecki, R.  Objectivity in a noisy photonic environment through quantum state information broadcasting.  {\em Phys. Rev. Lett.} {\bf 2014}, {\em 112}, 120402.

\bibitem{horodecki:15}
Horodecki, R.; Korbicz, J.K.; Horodecki, P. Quantum origins of objectivity. {\em Phys. Rev. A} {\bf 2015}, {\em 91}, 032122.

\bibitem{brandao:15}
Brand\~{a}o, F.G.S.L.; Piani, M.; Horodecki, P. Generic emergence of classical features in quantum Darwinism. {\em Nat. Commun.} {\bf 2015}, {\em 6}, 7908.

\bibitem{fields:10}
Fields, C.  Quantum Darwinism requires an extra-theoretical assumption of encoding redundancy. {\em Int. J. Theor. Phys.} {\bf 2010}, {\em 49}, 2523--2527.

\bibitem{fields:11}
Fields, C.  Classical system boundaries cannot be determined within quantum Darwinism. {\em Phys. Essays} {\bf 2011}, {\em 24}, 518--522.

\bibitem{schloss:06}
Schlosshauer, M. Experimental motivation and empirical consistency of minimal no-collapse quantum mechanics. {\em Ann. Phys.} {\bf 2006}, {\em 321}, 112–149.

\bibitem{everett:57}
Everett, H., III. ~``Relative state'' formulation of quantum mechanics. {\em Rev. Mod. Phys.} {\bf 1957},  {\em 29},  454--462.

\bibitem{ghirardi:86}
Ghirardi, G.C.; Rimini, A.; Weber, T. Unified dynamics for microscopic and macroscopic systems. {\em Phys.~Rev.~D} {\bf 1986}, {\em 34}, 470--491.

\bibitem{penrose:96}
Penrose, R. On gravity's role in quantum state reduction. {\em Gen. Relativ. Gravit.} {\bf 1996}, {\em 28}, 581--600.

\bibitem{weinberg:12}
Weinberg, S. Collapse of the state vector. {\em Phys. Rev. A} {\bf 2012}, {\em 85}, 062116.

\bibitem{jordan:10}
Jordan, T.F. Fundamental significance of tests that quantum dynamics is linear. {\em Phys. Rev.  A} {\bf 2010},  {\em 82}, 032103.

\bibitem{wiseman:15}
Wiseman, H. Quantum physics: Death by experiment for local realism. {\em Nature} {\bf 2015}, {\em 526}, 649--650.

\bibitem{swingle:12}
Swingle, B. Entanglement renormalization and holography. {\em Phys. Rev. D} {\bf 2012},  {\em 86}, 065007.

\bibitem{saini:15}
Saini, A.; Stojkovic, D. Radiation from a collapsing object is manifestly unitary.  {\em Phys. Rev. Lett.} {\bf 2015}, {\em 114},~111301.

\bibitem{susskind:16}
Susskind, L. Computational complexity and black hole horizons.  {\em Fortschr. Phys.} {\bf 2016},  {\em 64}, 24--43.

\bibitem{fields:12a}
Fields, C.  If physics is an information science, what is an observer?  {\em Information} {\bf 2012}, {\em 3}, 92--123.

\bibitem{fields:12b}
Fields, C.  A model-theoretic interpretation of environment-induced superselection.  {\em Int. J. Gen. Syst.} {\bf 2012}, {\em 41}, 847--859.

\bibitem{fields:14a}  
Fields, C.  On the Ollivier-Poulin-Zurek definition of objectivity.  {\em Axiomathes} {\bf 2014}, {\em 24}, 137--156.

\bibitem{fields:16a}
Fields, C.  Decompositional equivalence: A fundamental symmetry underlying quantum theory.  {\em Axiomathes} {\bf 2016}, {\em 26}, 279--311.

\bibitem{zanardi:01}
Zanardi, P.  Virtual quantum subsystems. {\em Phys. Rev. Lett.} {\bf 2001}, {\em 87}, 077901.

\bibitem{zanardi:04}
Zanardi, P.; Lidar, D.A.; Lloyd, S. Quantum tensor product structures are observable-induced. {\em Phys.~Rev.~Lett.} {\bf 2004}, {\em 92}, 060402.

\bibitem{dugic:06}
Dugi\'c, M.; Jekni\'c, J. What is ``system'': Some decoherence-theory arguments. {\em Int. J. Theor. Phys.} {\bf 2006}, {\em 45}, 2249--2259.

\bibitem{dugic:08}
Dugi\'c, M.; Jekni\'c - Dugi\'c, J. What is ``system'': The information-theoretic arguments. {\em Int. J. Theor. Phys.} {\bf 2008}, {\em 47}, 805--813.

\bibitem{torre:10}
De~la Torre, A.C.; Goyeneche, D.; Leitao, L. Entanglement for all quantum states. {\em Eur. J. Phys.} {\bf 2010}, {\em 31}, 325--332.

\bibitem{harshman:11}
Harshman, N.L.; Ranade, K.S. Observables can be tailored to change the entanglement of any pure state. {\em Phys. Rev. A} {\bf 2011}, {\em 84}, 012303.

\bibitem{thirring:11}
Thirring, W.; Bertlmann, R.A.; K{\"o}hler, P.; Narnhofer, H. Entanglement or separability: The choice of how to factorize the algebra of a density matrix. {\em Eur. Phys. J. D} {\bf 2011}, {\em 64}, 181--196.

\bibitem{dugic:12}
Dugi\'c, M.; Jekni\'c-Dugi\'c, J. Parallel decoherence in composite quantum systems. {\em Pramana} {\bf 2012}, {\em 79}, 199--209.

\bibitem{rovelli:96}
Rovelli, C.  Relational quantum mechanics. {\em Int. J. Theor. Phys.} {\bf 1996}, {\em 35}, 1637--1678.

\bibitem{moore:56}
Moore, E.F. Gedankenexperiments on sequential machines.  In {\em Autonoma Studies}; Shannon, C.W., McCarthy,~J.,~Eds.;   Princeton University Press:  Princeton, NJ, USA, 1956;  pp. 129--155.

\bibitem{popper:63}
Popper, K. {\em Conjectures and Refutations: The Growth of Scientific Knowledge};  Routledge \& Kegan Paul: London,~UK, 1963.

\bibitem{fuchs:14}
Fuchs, C.A.; Stacey, B.C.  Some negative remarks on operational approaches to quantum theory.  2014, preprint arxiv:1401.7254v1.

\bibitem{bell:64}
Bell, J.S.  On the Einstein-Podolsky-Rosen paradox.  {\em Physics} {\bf 1964}, {\em 1}, 195--200.

\bibitem{aspect:82}
Aspect, A.; Dalibard, J.; Roger, G. Experimental test of Bell's inequalities using time-varying analyzers. {\em Phys.~Rev. Lett.} {\bf 1982}, {\em 49}, 1804--1807.

\bibitem{fields:13a}
Fields, C.  Bell's theorem from Moore's theorem.  {\em Int. J. Gen. Syst.} {\bf 2013}, {\em 42}, 376--385.

\bibitem{kochen:67}
Kochen, S.; Specker, E.P.  The problem of hidden variables in quantum mechanics.  {\em J. Math.  Mech.} {\bf 1967}, {\em 17}, 59--87.

\bibitem{wootters:82}
Wootters, W.K.; Zurek, W.H.  A single quantum cannot be cloned.  {\em Nature} {\bf 1982}, {\em 299}, 802--803.

\bibitem{leifer:16}
Jennings, D.; Leifer, M.  No return to classical reality.  {\em Contemp. Phys.} {\bf 2016}, {\em 57}, 60--82.

\bibitem{peres:04}
Peres, A.; Terno, D.R.  Quantum information and relativity theory.  {\em Rev. Mod. Phys.} {\bf 2004}, {\em 76}, 93--123.

\bibitem{chitambar:14}
Chitambar, E.; Leung, D.; Man\v{c}inska, L.; Ozols, M.; Winter, A.  Everything you always wanted to know about LOCC (but were afraid to ask).  {\em Commun. Math. Phys.} {\bf 2014}, {\em 328}, 303--326.

\bibitem{bartlett:07}
Bartlett, S.D.; Rudolph, T.; Spekkens, R.W.  Reference frames, superselection rules, and quantum information.  {\em Rev. Mod. Phys.} {\bf 2007}, {\em 79}, 555--609.

\bibitem{roederer:05}
Roederer, J.G. {\em Information and Its Role in Nature};   Springer: Berlin, Germany, 2005.

\bibitem{roederer:16}
Roederer, J.G. Pragmatic information in biology and physics.  {\em Philos. Trans. R. Soc.  A} {\bf 2016}, {\em 374},  20150152.

\bibitem{asaro:08}
Asaro, P.M.  From mechanisms of adaptation to intelligence amplifiers: The philosophy of W. Ross Ashby.  In~{\em The Mechanical Mind in History}; Husbands, P.,  Holland, O.,  Wheeler, M.,  Eds.; MIT/Bradford:  Cambridge, MA, USA, 2008.

\bibitem{bekenstein:73}
Bekenstein, J.D.  Black holes and entropy.  {\em Phys. Rev. D} {\bf 1973}, {\em 7}, 2333--2346.

\bibitem{harlow:13}
Harlow, D.; Hayden, P.  Quantum computing vs. firewalls.  {\em J. High Energy Phys.} {\bf 2013}, {\em 2013}, 85.

\bibitem{conway:06}
Conway, J.; Kochen, S.  The free will theorem.  {\em Found.  Phys.} {\bf 2006}, {\em 36}, 1441--1473.

\bibitem{fields:13b}
Fields, C.  A whole box of Pandoras: Systems, boundaries and free will in quantum theory.  {\em J. Exp. Theor.  Artif.~Intell.} {\bf 2013}, {\em 25}, 291--302.

\bibitem{wang:94}
Wang, Q.; Schoenlein, R.W.; Peteanu, L.A.; Mathies, R.A.; Shank, C.V.  Vibrationally coherent
photochemistry in the femtosecond primary event of vision.  {\em Science} {\bf 1994}, {\em 266}, 422--424.


\bibitem{wheeler:90}
Wheeler, J.A.  Information, physics, quantum: The search for links.  In  {\em Complexity, Entropy, and the Physics of Information}; Zurek, W. H.,  Ed.;   Westview:  Boulder, CO, USA, 1990;  pp. 3--28.

\bibitem{clifton:03}
Clifton, R.; Bub, J.; Halvorson, H. Characterizing quantum theory in terms of information-theoretic
constraints. {\em Found.  Phys.} {\bf 2003}, {\em 33}, 1561--1591.

\bibitem{ariano:11}
D'Ariano, G.M.  Physics as Information Processing.   In Proceedings of the Physics Education Research Conference, Omaha, NE, USA,  3--4 August  2011; Volume 1327,  pp. 7--18.

\bibitem{chiribella:11}
Chiribella, G.; D’Ariano, G.M.; Perinotti,  P.  Informational derivation of quantum theory. {\em Phys.
Rev. A} {\bf 2011}, {\em 84}, 012311.

\bibitem{chiribella:12}
Chiribella, G.; D’Ariano, G.M.; Perinotti, P.  Quantum theory, namely the pure and reversible theory of information.  {\em Entropy} {\bf 2012}, {\em 14}, 1877--1893.

\bibitem{hardy:15}
Hardy, L.  Reconstructing quantum theory.  {\em Fund. Theor.  Phys.} {\bf 2015}, {\em 181}, 223--248.

\bibitem{muller:15}
M\"{u}ller, M.P.; Masanes, L.  Information-theoretic postulates for quantum theory.  {\em Fund. Theor.  Phys.} {\bf 2015}, {\em 181}, 139--170.

\bibitem{knuth:14}
Knuth, K.H.  Information-based physics: An observer-centric foundation.  {\em Contemp. Phys.} {\bf 2014}, {\em 55}, 12--32.

\bibitem{deutsch:15}
Deutsch, D.; Marletto, C.  Constructor theory of information.  {\em Proc. R. Soc. A} {\bf 2015}, {\em 471}, 20140540.

\bibitem{hohn:15}
H\"{o}hn, P.A.; Wever, C.S.P.  Quantum theory from questions.  2015, preprint arxiv:1511.01120v2.

\bibitem{grinbaum:15}
Grinbaum, A.  How device-independent approaches change the meaning of physical theory.  2015, preprint arxiv:1512.01035v3.

\bibitem{zimmer:10}
Zimmer, H.D.; Ecker, U.K.D.  Remembering perceptual features unequally bound in object and episodic tokens: Neural mechanisms and their electrophysiological correlates.  {\em Neurosci. Biobehav. Rev.} {\bf 2010}, {\em 34}, 1066--1079.

\bibitem{fields:12c}
Fields, C.  The very same thing: Extending the object token concept to incorporate causal constraints on individual identity.  {\em Adv. Cognit. Psychol.}  {\bf 2012}, {\em 8}, 234--247.

\bibitem{harrison:73}
Harrison, E.R.  Standard model of the early universe.  {\em Annu. Rev. Astron. Astrophys.} {\bf 1973}, {\em  11}, 155--186.

\bibitem{fields:16b}
Fields, C.  Visual re-identification of individual objects: A core problem for organisms and AI.  {\em Cognit. Process.} {\bf 2016}, {\em 17}, 1--13.

\bibitem{arkani:14}
Arkani-Hamed, N.; Trnka, J.  The amplituhedron.  {\em J. High-Energy Phys.} {\bf 2014}, {\em 10}, 030.

\bibitem{krechmer:16}
Krechmer, K.  Relational measurements and uncertainty.  {\em Measurement} {\bf 2016},  doi:10.1016/j.measurement.2016.06.058.

\bibitem{friston:10}
Friston, K.  The free-energy principle: A unified brain theory?  {\em Nat.  Rev. Neurosci.} {\bf 2010}, {\em 11}, 127--138.

\bibitem{kauffman:03}
Kauffman, L.  Eigenforms---Objects as tokens for eigenbehaviors.  {\em Cybern.  Hum.  Knowing} {\bf 2003}, {\em 10}, 73--90.

\bibitem{kauffman:05}
Kauffman, L.  EigenForm.  {\em Kybernetes} {\bf 2005}, {\em 34}, 129--150.

\bibitem{dietrich:15}
Dietrich, E.; Fields, C.  Science generates limit paradoxes.  {\em Axiomathes} {\bf 2015}, {\em 25}, 409--432.

\bibitem{witt:22}
Wittgenstein, L.  {\em Tractatus Logico-Philosophicus}.  Kegan Paul, Trench, Trubner:  London, UK, 1922.

\bibitem{tarski:44}
Tarski, A.  The semantic conception of truth and the foundations of semantics.  {\em Philos. Phenomenol.  Res.} {\bf 1944}, {\em 4}, 341--376.

\bibitem{turing:36}
Turing, A.R.  On computable numbers, with an application to the {\em Entscheidungsproblem}.  {\em Proc. Lond. Math. Soc.} {\bf 1936}, {\em 442}, 230--265.

\bibitem{quine:48}
Quine, W.V.O.  On what there is.  {\em Rev. Metaphys.} {\bf 1948}, {\em 2}, 21--38.

\bibitem{tan:76} 
Tanenbaum, A.S.  {\em Structured Computer Organization};  Prentice Hall:  Upper Saddle River, NJ, USA,  1976.

\bibitem{smith:05}
Smith, J.E.; Nair, R.  The architecture of virtual machines.  {\em IEEE Comput.} {\bf 2005},  {\em 38}, 32--38.

\bibitem{partridge:10}
Partridge, D.  {\em The Seductive Computer};   Springer: London, UK, 2010.

\bibitem{chomsky:59}
Chomsky, N.  Review of  B. F. Skinner, {\em Verbal Behavior}.  {\em Language}  {\bf 1959}, {\em 35}, 26--58.

\bibitem{harnad:90}
Harnad, S.  The symbol grounding problem.  {\em Physica D} {\bf 1990}, {\em 42}, 335--346.

\bibitem{taddeo:05}
Taddeo, M.; Floridi, L.  Solving the symbol grounding problem: A critical review of fifteen years
of research.  {\em J. Exp. Theor. Artif. Intell.} {\bf 2005}, {\em 17}, 419--445.

\bibitem{fields:14b}
Fields, C.  Equivalence of the symbol grounding and quantum system identification problems.  {\em Information} {\bf 2014}, {\em 5}, 172--189.

\bibitem{fodor:80}
Fodor, J.A.  Methodological solipsism considered as a research strategy in cognitive science.  {\em Behav. Brain Sci.} {\bf 1980}, {\em 3}, 63--73.

\bibitem{clark:98}
Clark, A; Chalmers, D.  The extended mind.  {\em Analysis} {\bf 1998}, {\em 58}, 7--19.

\bibitem{anderson:03}
Anderson, M.L.  Embodied cognition: A field guide.  {\em Artif. Intell.} {\bf 2003}, {\em 149}, 91--130.

\bibitem{cangelosi:15}
Cangelosi, A.; Schlesinger, M.  {\em Developmental Robotics: From Babies to Robots};  MIT Press: Cambridge, MA,  USA, 2015.

\bibitem{hommel:04}
Hommel, B.  Event files: Feature binding in and across perception and action.  {\em Trends Cognit. Sci.} {\bf 2004}, {\em 8}, 494--500.

\bibitem{eichenbaum:07}
Eichenbaum, H.; Yonelinas, A.R.; Ranganath, C.  The medial temporal lobe and recognition memory.  {\em Annu.~Rev. Neurosci.} {\bf 2007}, {\em 30}, 123--152.

\bibitem{cubek:15}
Cubek, R.; Ertel, W.; Palm, G.  A critical review on the symbol grounding problem as an issue of autonomous agents.  {\em Lect.  Notes  Comput. Sci.} {\bf 2015}, {\em 9324}, 256--263.

\bibitem{mark:10}
Mark, J.T.; Marion, B.B.; Hoffman, D.D.  Natural selection and veridical perceptions.  {\em J. Theor. Biol.} {\bf 2010}, {\em 266}, 504--515.

\bibitem{hoffman:15}
Hoffman, D.D.; Singh, M.; Prakash, C.  The interface theory of perception.  {\em Psychon. Bull. Rev.} {\bf 2015}, {\em 22}, 1480--1506.

\bibitem{hoffman:14}
Hoffman, D.D.; Prakash, C.  Objects of consciousness.  {\em Front. Psychol.} {\bf 2014}, {\em 5}, 577.

\bibitem{friston:11}
Friston, K.J.  Functional and effective connectivity: A review.  {\em Brain Connect.} {\bf 2011}, {\em 1}, 13--36.

\bibitem{friston:15}
Friston, K.; Levin, M.; Sengupta, B.; Pezzulo, G.  Knowing one’s place: A free-energy approach to pattern regulation.  {\em J. R. Soc. Interface} {\bf 2015}, {\em 12}, 20141383.

\bibitem{friston:13}
Friston, K.  Life as we know it.  {\em J. R. Soc. Interface} {\bf 2013},  {\em 10}, 20130475.

\bibitem{lawson:14}
Lawson, R.P.; Rees, G.; Friston, K.J. An aberrant precision account of autism. {\em Front. Hum. Neurosci.} {\bf 2014},~{\em 8},~302.

\bibitem{bostom:03}
Bostom, N.  Are we living in a computer simulation?  {\em Philos. Quart.} {\bf 2003}, {\em 53}, 243--255.


\end{thebibliography}
\end{document}